\documentclass{article}

\usepackage{arxiv}

\usepackage[utf8]{inputenc} 
\usepackage[T1]{fontenc}    
\usepackage{hyperref}       
\usepackage{url}            
\usepackage{booktabs}       
\usepackage{amsfonts}       
\usepackage{nicefrac}       
\usepackage{microtype}      
\usepackage{lipsum}
\usepackage{graphicx}
\usepackage{enumitem}
\usepackage{gensymb}
\usepackage{amsmath}
\usepackage{amssymb}
\usepackage{float}

\usepackage{mathtools}
\DeclarePairedDelimiter{\nint}\lfloor\rceil

\usepackage[ruled,linesnumbered]{algorithm2e}
\makeatletter
\newcommand{\nosemic}{\renewcommand{\@endalgocfline}{\relax}}
\newcommand{\dosemic}{\renewcommand{\@endalgocfline}{\algocf@endline}}
\newcommand{\pushline}{\Indp}
\newcommand{\popline}{\Indm\dosemic}
\let\oldnl\nl
\newcommand{\nonl}{\renewcommand{\nl}{\let\nl\oldnl}}
\makeatother

\usepackage[style=numeric-comp, sorting=none]{biblatex}
\title{Particle tracking velocimetry in liquid gallium flow about a cylindrical obstacle}

\author{
  Mihails Birjukovs\\
  Institute of Numerical Modelling\\
  University of Latvia (UL)\\
  Riga, Latvia, Jelgavas 3, 1004 \\
  \texttt{mihails.birjukovs@lu.lv} \\
   \And
  Peteris Zvejnieks\\
  Institute of Numerical Modelling\\
  University of Latvia (UL)\\
  Riga, Latvia, Jelgavas 3, 1004 \\
  \texttt{peteris.zvejnieks@lu.lv} \\
   \And
  Tobias Lappan\\
  Helmholtz-Zentrum Dresden-Rossendorf (HZDR)\\
  Department of Magnetohydrodynamics\\
  Department of Transport Processes at Interfaces\\
  Bautzner Landstraße 400, 01328 Dresden, Germany \\
  \And
  Martins Sarma\\
  Helmholtz-Zentrum Dresden-Rossendorf (HZDR)\\
  Department of Magnetohydrodynamics\\
  Bautzner Landstraße 400, 01328 Dresden, Germany \\
  \And
  Sascha Heitkam\\
  Helmholtz-Zentrum Dresden-Rossendorf (HZDR)\\
  Department of Transport Processes at Interfaces\\
  Bautzner Landstraße 400, 01328 Dresden, Germany \\
  Technische Universität Dresden \\
  Institute of Process Engineering and Environmental Technology\\
  01062 Dresden, Germany
  \And
  Pavel Trtik \\
  Laboratory for Neutron Scattering and Imaging\\
  Paul Scherrer Institut\\
  Villigen, Switzerland, Forschungsstrasse 111, 5232 \\
  \texttt{pavel.trtik@psi.ch} \\
  \And
  David Mannes \\
  Laboratory for Neutron Scattering and Imaging\\
  Paul Scherrer Institut\\
  Villigen, Switzerland, Forschungsstrasse 111, 5232 \\
  \And
    Sven Eckert\\
  Helmholtz-Zentrum Dresden-Rossendorf (HZDR)\\
  Department of Magnetohydrodynamics\\
  Bautzner Landstraße 400, 01328 Dresden, Germany \\
   \And
    Andris Jakovics\\
  Institute of Numerical Modelling\\
  University of Latvia (UL)\\
  Riga, Latvia, Jelgavas 3, 1004
  }

\begin{document}
\maketitle

\clearpage

\begin{abstract}
This paper demonstrates particle tracking velocimetry performed for a model system wherein particle-laden liquid metal flow about a cylindrical obstacle was studied. We present the image processing methodology developed for particle detection in images with disparate and often low signal- and contrast-to-noise ratios, and the application of our MHT-X tracing algorithm for particle trajectory reconstruction within and about the wake flow of the obstacle. Preliminary results indicate that the utilized methods enable consistent particle detection and recovery of long, representative particle trajectories with high confidence. However, we also underline the necessity of implementing a more advanced particle position extrapolation approach for increased tracking accuracy. We also show that the utilized particles exhibit the Stokes number range that suggests good flow tracking accuracy, and that the particle time scales extracted from reconstructed trajectories are consistent with expectations based \textit{a priori} estimates.
\end{abstract}


\section{Introduction}

Liquid metal stirring, continuous casting, chemical reactors, etc. in many cases involve bubble flow in liquid metal. Some of these processes are, or potentially can be, controlled using applied magnetic field (MF) \cite{baakeNeutronRadiographyVisualization2017, casting-euler-musig, embr-experiment, embr-experiment-2, limmcast, embr-visualized, embr-cift, birjukovsPhaseBoundaryDynamics2020}. To do this, however, bubble flow physics without and with applied magnetic field must be well understood.

Single bubble magnetohydrodynamic (MHD) flow, and also without applied MF, has been extensively studied by means of ultrasound Doppler velocimetry (UDV) \cite{zhang-thesis, udv-review-article, udv-longitudinal-field, udv-transverse-field}, ultrasound transit time technique \cite{uttt-path-instability, uttt-x-ray-single-bubble}, X-ray imaging \cite{uttt-x-ray-single-bubble} and numerical simulations \cite{hzdr-ibm-bubbles-thesis, dns-longitudinal-field, imb-transverse-field, zhang-mf-vertical, zhang-mf-simulations, gaudlitz-shape-wake-variations-1-bubble, hele-shaw-bubbles-vof, hele-shaw-bubbles-experiment}, and many of its characteristics and mechanisms are presently sufficiently clear \cite{prl-path-instability, natcomms-shape-dynamics, spiral-to-zigzag-explained, shape-and-wake-simulations, spiral-to-zigzag-explained-2, in-depth-study-of-ellipsoid-kinematics}. However, many aspects of bubble collective dynamics, especially in presence of MF, are not properly understood or have not been studied at all \cite{x-ray-bubble-breakup, x-ray-bubble-coalescence, optics-collective-dynamics, 2021-review-article-bubbles-in-liquid-metal}. This makes it impossible to significantly improve effective models for bubble flow (Euler-Euler and Lagrangian) and the above mentioned industrial processes without insights into how bubbles interact in MHD flow (or even without applied MF) \cite{casting-euler-les, casting-euler-musig, casting-lagrange-bubbles, casting-new-collective-dynamics-models, taborda-les-euler-lagrange}.

In addition, bubble interaction with particles is of interest in metal purification \cite{metal-strring-1, metal-strring-2, metal-strring-3, metal-strring-4, metal-strring-5} and froth flotation \cite{flotation-book-1, flotation-book-2, sommer-4d-ptv}. It is also known that bubble wake flow is what primarily determines bubble trajectories in absence of perturbations from other bubbles \cite{uttt-path-instability, prl-path-instability, spiral-to-zigzag-explained, spiral-to-zigzag-explained-2}. Despite this, there is very little experimental work where bubble wakes or bubble/particle interactions are directly visualized in liquid metal \cite{lappan2020a}. UDV has been applied to bubble wake flow characterization \cite{zhang-thesis, udv-wake-flow-structure}, and the potential of flow analysis via particle tracking in liquid metal using positron emission particle tracking (PEPT) has been explored \cite{sommer-pept, pept-1, pept-2, pept-3, pept-4}.

Through recent studies and the advent of dynamic X-ray and neutron radiography of two-phase liquid metal flow \cite{megumi-x-rays, saito-neutrons-1, saito-neutrons-2, lappan2020a, neutrons-particles-stirrer-scepanskis, neutrons-particles-stirrer-scepanskis-2, neutrons-simulations-stirrer-valters}, however, fundamental investigation of bubble flow systems mimicking industrially relevant flow conditions is underway \cite{hzdr-ibm-bubbles-thesis, birjukovsArgonBubbleFlow2020, birjukovsPhaseBoundaryDynamics2020, birjukovs2021resolving, baakeNeutronRadiographyVisualization2017, x-ray-bubble-chain-simulate, x-ray-prime-code, x-ray-bubble-breakup, x-ray-bubble-coalescence, x-ray-validation, megumi-x-rays}. Such systems are typically rectangular vessels filled with gallium \cite{birjukovsArgonBubbleFlow2020, birjukovsPhaseBoundaryDynamics2020, baakeNeutronRadiographyVisualization2017} or an eutectic gallium-indium-tin alloy \cite{x-ray-bubble-chain-simulate, x-ray-bubble-breakup, x-ray-bubble-coalescence, x-ray-validation, megumi-x-rays}, where bubbles are introduced via nozzles which usually are horizontal \cite{birjukovsArgonBubbleFlow2020,birjukovsPhaseBoundaryDynamics2020} or vertical \cite{baakeNeutronRadiographyVisualization2017, x-ray-bubble-chain-simulate, x-ray-bubble-breakup, x-ray-bubble-coalescence, x-ray-validation} at the bottom of the vessel, or top-submerged vertical \cite{megumi-x-rays}. Recently, dynamic mode decomposition has been applied to the output of the MHD bubble chain flow simulation to study both large-scale flow structures and bubble wake flow in the bubble reference frame \cite{klevs2021dynamic}. The developed approach could be applied to experimental data as well.

It was proposed some time ago that neutron radiography could also be used to directly observe wake flow of bodies within optically opaque systems \cite{neutrons-fluid-flow-visuals}. The first such benchmark study in the context of liquid metal flow with dispersed particles was done by Lappan et al. where gadolinium oxide particle flow about a cylindrical obstacle in a thin liquid metal channel was imaged dynamically with sufficient temporal resolution using high cold neutron flux \cite{lappan2020a, x-ray-neutron-experiments-book}. The imaged particle flow was investigated with particle image velocimetry (PIV) and the wake flow velocity field was measured and visualized. The idea was that such model experiments could be used to represent and study wake flow behind single bubbles ascending through liquid metal. In the present paper, we take this a step further by performing particle tracking velocimetry (PTV). Unlike PIV which uses image feature correlations for velocimetry, PTV is explicit particle tracking where particles are treated as point-like bodies. Thus, PTV allows to quantify the dynamics of individual particles explicitly and enables wake flow analysis at finer length scales.

Particle tracking in liquid metals using neutron radiography is an important problem that can be solved using high-resolution neutron radiography. However, only a very limited number of papers address image processing required to successfully extract physically meaningful information from the acquired image. Notably, Heitkam et al. performed particle detection and tracking in froth using neutron imaging utilizing a particle-mask correlation approach \cite{heitkam-particles-froth-2019}. An original approach for detecting particles and tracking particle flow in the presence of bubbles was demonstrated by Sommer et al., although not in the context of liquid metal \cite{sommer-4d-ptv}. An approach that is very promising for particle detection within flow with a high particle number density was developed by Anders et al. for optical measurements, but could potentially be generalized \cite{sten-spectral-random-masking, sten-srm-velocity-temp-measurements}. However, the latter two do not seem to be readily applicable to low signal-to-noise ratio (SNR) images typically associated with high frame rate neutron imaging, and the former would be hard to generalize due to its reliance on preset particle masks. In addition, it is advantageous to combine a more noise-resilient image processing approach with a more general method for particle tracking based on detections.

We have therefore developed an image processing methodology that reliably extracts particles of various sizes and visibility from sequences of images with a low SNR and spatially correlated noise. We have then input the set of particle locations detected over time to our previously developed  MHT-X object tracking algorithm \cite{zvejnieks2021mhtx} based on a general and robust framework of multiple hypothesis tracking (MHT), and were able to reconstruct particle trajectories within and about the wake flow zone behind the cylindrical obstacle. Later we also point out the current limitations of our methods and propose future improvements that could increase the PTV quality.

\section{The experiment}
\label{sec:experiment}

\begin{figure}[h]
			\centering
			\includegraphics[width=0.8\linewidth]{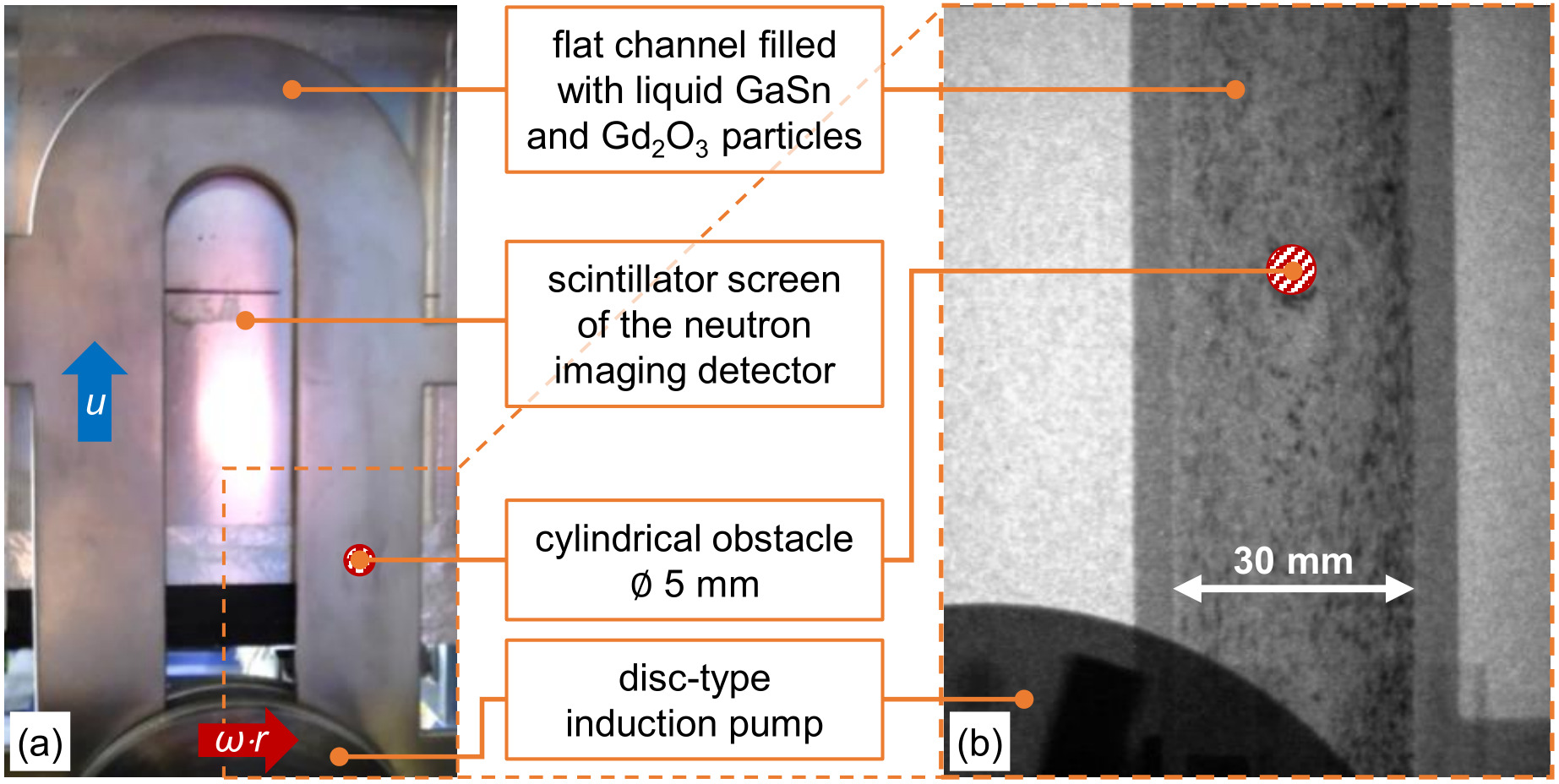}
			\caption{(a) A photograph of the experimental setup and (b) a neutron radiograph of particle flow in liquid metal within the channel about the cylindrical obstacle \cite{lappan2020a}.}
			\label{fig:experimental-setup}
	\end{figure}

\begin{figure}[h]
			\centering
			\includegraphics[width=1\linewidth]{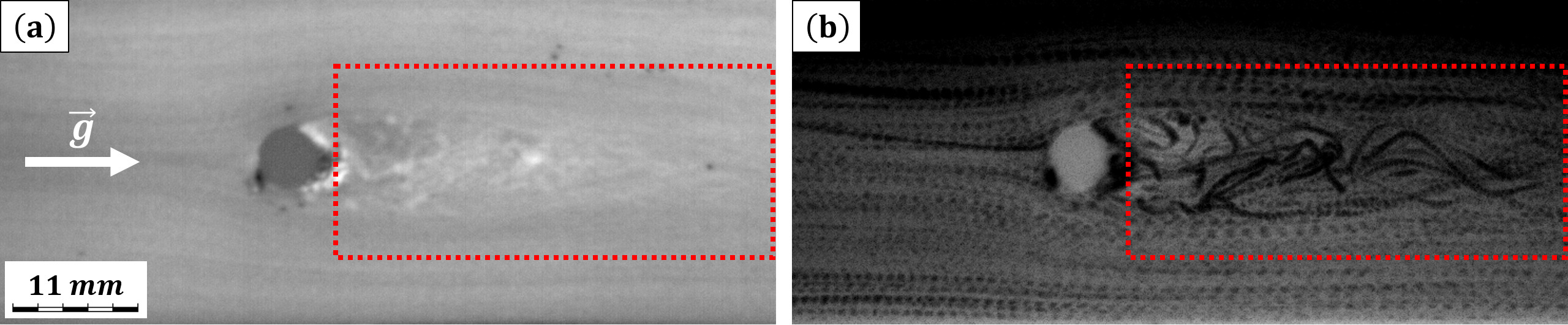}
			\caption{Pixel-wise (a) standard deviation and (b) minima of luminance values within the imaged flow channel over all captured frames. The region of interest is indicated with a red dashed frame. Note the cylindrical obstacle in both figures. The images shown here were rotated 90 degrees left with respect to the originals (Figure \ref{fig:experimental-setup}b) -- as such, here the originally downward flow is directed from left to right. The white arrow in (a) indicates the gravitational acceleration $\vec{g}$.}
			\label{fig:fov-stdv-min-projections}
	\end{figure}

	This section gives a brief overview of the experiment and the dynamic neutron radiography setup used for flow imaging. The experimental setup is shown in Figure \ref{fig:experimental-setup} and more details can be found in our previous paper \cite{lappan2020a}. Metal and particle flow about the cylinder shown in Figures \ref{fig:experimental-setup}b and \ref{fig:fov-stdv-min-projections} is a model system for a single bubble rising in the liquid metal, where the aim is to detect particles and reconstruct their trajectories to analyze the wake flow. Given that neutron transmission imaging yields particle projections, it is desirable to avoid three-dimensional motion, so this experiment is performed for a quasi two-dimensional geometry. Consequently, the bubble is represented by a cylindrical obstacle instead of a spherical one. The cylinder with a $5~mm$ diameter is made from stainless steel (X5CrNi18-10) and is centered and fixed in the straight section of the flow channel. The boundary of such an obstacle should obey the no-slip condition. However, note that even if this is not exactly the case, the differences in the wake shape are not expected to be significant \cite{particle-boundary-conditions}. The flow channel has a uniform $30~ mm \times 3~mm$ rectangular cross-section -- flow was imaged through the $3~mm$ dimension. To generate continuous liquid metal flow about the cylinder, the flow channel was designed as a closed loop. The loop is made from the same material as the cylindrical obstacle. Liquid metal flow is driven by a disc-type electromagnetic induction pump equipped with permanent magnets \cite{lappan2020a}.

	Flow measurements were performed at room temperature without additional heating using a low melting point gallium-tin alloy. Compared to pure gallium with the $30 \degree C$ melting point, the binary alloy with a $0.07$ tin mass fraction is liquid at a slightly lower temperature of $25 \degree C$ \cite{anderson1992}. In this experiment, we opted for small particles made of gadolinium oxide. Gadolinium has an extremely high neutron attenuation coefficient $\mu_{Gd} = 1.5 \cdot 10^3~ {cm}^{-1}$ compared to gallium with $\mu_{Ga} = 0.5~ {cm}^{-1}$ and tin $\mu_{Sn} = 0.2 ~ {cm}^{-1}$ \cite{sears1992, PSI_LinAttCoeffTher}. Gadolinium oxide particles with a $d_\text{p} \in (0.3;0.5)~mm$ diameter have been shown to provide sufficient image contrast for dynamic neutron imaging with a short image exposure time \cite{lappan2020a}. Note that given the $3~mm$ thickness in the neutron flux direction the gallium tin alloy is rather transparent ($\sim 87 \%$ transmission) to neutron radiation. Gadolinium oxide also has a lower paramagnetic susceptibility than gadolinium~\cite{lide2019, martienssen2005}. Gadolinium particles have been found to be inapplicable to experiments like the one outlined here: they are strongly attracted by the magnetic field of the electromagnetic induction pump and thus tend to rapidly clog the flow channel. Using gadolinium oxide particles instead, liquid metal flow can be driven continuously without interruptions for cleanup. Other metal and particle properties relevant for subsequent analysis include: $GaSn$ density $\rho_0 = 6160~ kg/m^3$; $GaSn$ viscosity $\mu_0 = 2.1~ mPa \cdot s$; particle density $\rho_\text{p} = 7410~ kg/m^3$.
	
	Neutron radiography allows to visualize the gadolinium oxide particles and track their motion within the optically opaque liquid metal. Imaging was carried out with cold neutrons at the ICON beamline \cite{kaestner2011} of the Swiss spallation neutron source SINQ \cite{blau2009}. During the measurements, the neutron source was operated at a constant proton beam current of $1.3~ mA$. The neutron flux $\phi$ is an important parameter for dynamic neutron imaging and significantly depends on the neutron beam aperture $D$. Most measurements were performed with $D = 40~mm$, providing $\phi = 5 \cdot 10^7~ n \cdot cm^{-2} s^{-1}$. For a few measurement runs, the neutron beam aperture was doubled to $D = 80~mm$, which increases the neutron flux to $\phi = 1.8 \cdot 10^8~ n \cdot cm^{-2} s^{-1}$ (by a factor of $>3$) \cite{kaestner2011}. The neutron beam aperture, the flow experiment and the neutron scintillator screen of the imaging detector were aligned at fixed positions. The distance $L_{D}$ between the beam aperture and the scintillator was $\sim 6.9~m$, yielding the collimation ratio $L_{D}/D \sim 172$ for $D = 40~mm$. The distance $L_{exp}$ between the scintillator and the center plane of the liquid metal loop was $50~mm$ \cite{lappan2020a}. The resulting geometrical unsharpness is $L_{exp}/(L_{D}/D) \sim 0.3~mm$ which is about the size of the smallest gadolinium oxide particles. 
	
	We used a ${}^{6}$LiF:ZnS scintillator with a $200~\mu m$ thickness and a $150 \times 150~ mm$ observable area. The light emitted by the scintillator was acquired by a sCMOS camera (\textit{Hamamatsu ORCA Flash 4.0}; photographic objective: \textit{Nikon AF-S Nikkor 50mm 1:1.4G}). The camera's field of view was set to $100~mm \times 100~ mm$. The camera has a pixel array of $2048 \times 2048$ in total. Applying $2 \times 2$ pixel binning (average) for image noise reduction, the effective pixel array was reduced to $1024 \times 1024$. These camera settings result in a $10~px/mm$ spatial resolution. We chose a $10~ ms$ image exposure time equivalent 100 frames per second required to capture individual particles moving in the liquid metal flow.

\section{Particle detection}

\subsection{Image characterization \& considerations for image processing}

The region of interest where particles must be detected and tracked \textit{explicitly} is the wake flow area highlighted in Figure \ref{fig:fov-stdv-min-projections}: note that this is where the flow is, as expected, the most disordered (Figure \ref{fig:fov-stdv-min-projections}a) and particle tracks, which can be visualized using the minimum luminance projection in time (Figure \ref{fig:fov-stdv-min-projections}b), are strongly affected by the turbulent wake as opposed to very smooth trajectories to the sides of the wake flow zone. The analyzed field of view (FOV) was equal to $408 \times 161$ pixels (16-bit single-channel images) corresponding to $37.8~ mm \times 14.9~ mm$.

It was shown in \cite{heitkam-particles-froth-2019} that the images exhibit correlated noise in the form of grain-like structures with $\sim 3$-pixel sizes -- this is a considerable fraction of a typical particle size in images and thus images may contain "phantom" particles. Particle projections visible due to neutron transmission contrast have strongly varying sizes and signal- and contrast-to-noise ratios (SNR and CNR, respectively) that also change over time as particles travel through the FOV. In addition, the recorded image sequence exhibits a pronounced global luminance non-uniformity, which can be seen in Figure \ref{fig:fov-stdv-min-projections}b. Given these factors, it was decided to build the image processing procedure around a local filter applied to interrogation windows (IWs, not to be confused with PIV terminology) taken from the images. Due to the very high area density of particles in the images, instead of utilizing a combination of segment estimation via a global filter with subsequent iterative refinement using targeted local filtering as in \cite{birjukovs2021resolving}, it was decided to scan images entirely with partially overlapping IWs.

\subsection{The algorithm}

\begin{figure}[h]
			\centering
			\includegraphics[width=0.95\linewidth]{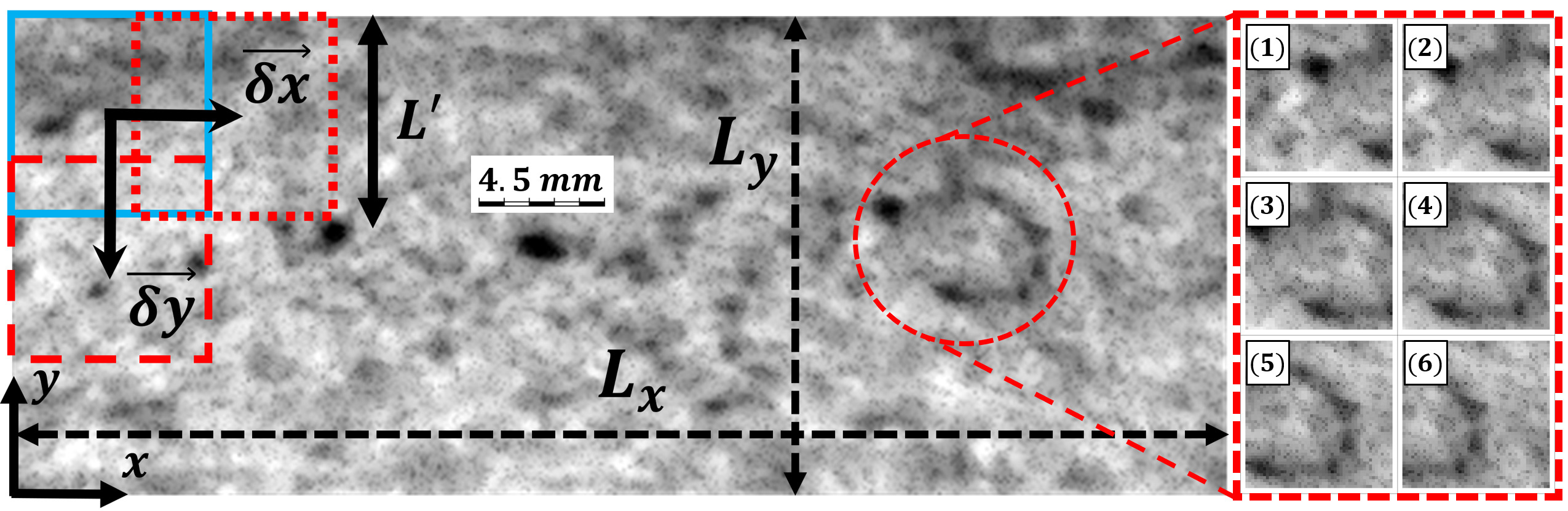}
			\caption{A schematic representation of the IW sweep for the FOV: the initial IW (light blue) and shifted IWs (red) in the $x>0$ (dotted) and $y<0$ (dashed) directions. An example a horizontal scan in the $x>0$ direction for an area indicated with a red dashed circle is shown in sub-figures 1-6.}
			\label{fig:iw-sweep}
	\end{figure}

The idea of sweeping images with partially overlapping IWs is illustrated in Figure \ref{fig:iw-sweep}. An initial square IW with a side length $L'$ is fitted into the upper-left corner of the image. A set of IWs is then generated from the initial IW by creating an IW position lattice with horizontal/vertical stepping $\delta x$ and $\delta y$, respectively, where the latter are a fraction of $L'$. Thus, a set of $n_x$ by $n_y$ partially overlapping IWs is created where the number of steps in each direction $n_k$ is determined by

\begin{equation}
    n_k = \nint{ \frac{L_k - L'}{\delta x_k} }, ~~ k = x,y
    \label{eq:iw-steps}
\end{equation}
where $L_k$ are the FOV dimensions. The $L_k - L'$ term ensures that the IWs are not excessively out of FOV bounds, since out-of-bounds parts of IWs are cropped, and overly cropped IW images do not provide enough meaningful information for local filtering. IW bounds $\text{IW}_{km}$ and centroids $\vec{r}_{km}$ are given by

\begin{equation}
    \text{IW}_{km}: 
    \bigg \{ ~
    \left[ 1 + k \cdot \delta x ;~ 
    L' + k \cdot \delta x \right] , 
    \left[ 1 + m \cdot \delta y ;~ 
    L' + m \cdot \delta y \right] 
    ~ \bigg \} ; ~~ k \in [0, n_x],~ m \in [0, n_y]
\label{eq:iw-bounds}
\end{equation}

\begin{equation}
    \vec{r}_{km} = \vec{e}_x \cdot \left( (1+L')/2 + k \cdot \delta x \right) +
    \vec{e}_y \cdot \left( (1+L')/2 + m \cdot \delta y \right); 
    ~~ k \in [0, n_x],~ m \in [0, n_y]
\label{eq:iw-centroids}
\end{equation}

All image processing operations are performed in \textit{Wolfram Mathematica}. The general framework for image processing is as shown in Algorithm \ref{alg:image-proc-framework}. Note that in this case we use the following parameters: $L' = 50$ (in pixels), $\delta x = \delta y = 10$. The values for $\delta x_k$ and $L'$ were chosen such that $\delta x_k$ is greater than the characteristic particle size, but otherwise a small fraction of $L'$; the latter was set to roughly match the expected scale of particle clusters seen in the FOV images. This is to ensure sufficient redundant detection for particles -- we observe that the selected values are optimal for our case in terms of detection accuracy. However, smaller $\delta x_k$ make such an approach more computationally expensive. For instance, with the current settings, a $37 \times 12$ grid of IWs is generated for a total of 444 local images. The degree of redundancy for IWs is determined by $1 - \delta x_k / L'$ which in this case is $80\%$. This effectively inflates the amount of data (total image area) by a factor of $\sim 16.9$. Therefore, one should take care to optimize the underlying image processing code in terms of memory utilization and parallelize as many of its elements as possible.

\begin{algorithm}

    \nonl \textbf{Input:} A sequence of normalized images (pixel luminance rescaled to $[0;1]$) with subtracted mean dark current
    
    \nonl \underline{\textit{IW generation}}
    
    \pushline Generate an $n_x \times n_y$ grid of IW positions (\ref{eq:iw-centroids}) based on (\ref{eq:iw-steps}), and the chosen $L'$, $\delta x$ and $\delta y$
    
    Disassemble images into their projections onto IWs (\ref{eq:iw-bounds})
    
    \nonl \popline \underline{\textit{Particle detection in IWs}}
    
    \pushline Normalize the IW images
    
    Local filtering (Algorithm \ref{alg:local-filter})
    
    Particle segmentation (Algorithm \ref{alg:local-segmentation})
    
    Luminance map-based false positive filtering (Algorithm \ref{alg:false-positive-luminance-filter})
    
    \nonl \popline \underline{\textit{Assembly of global detection masks}}
    
    \pushline Map the false positive-filtered IW particle masks onto the full FOV using (\ref{eq:iw-centroids})
    
    Sum the particle masks over the FOV
    
    Minimum area thresholding
    
    Morphological opening
    
    \popline Compute centroids for the resulting segments

    \nonl \textbf{Output:} Centroids for particles detected in every FOV image
    
\caption{Image processing framework}
\label{alg:image-proc-framework}
\end{algorithm}

\begin{algorithm}

    \nonl \textbf{Input:} A normalized IW image
    
    Invert the image luminance map
    
    Non-local means masking (NMM)
    
    Soft color tone map masking (SCTMM)
    
    Non-local means (NM) filtering
    
    Mean filtering

    \nonl \textbf{Output:} A filtered IW image
    
\caption{Local (IW) filtering}
\label{alg:local-filter}
\end{algorithm}

\begin{figure}[h]
			\centering
			\includegraphics[width=0.9\linewidth]{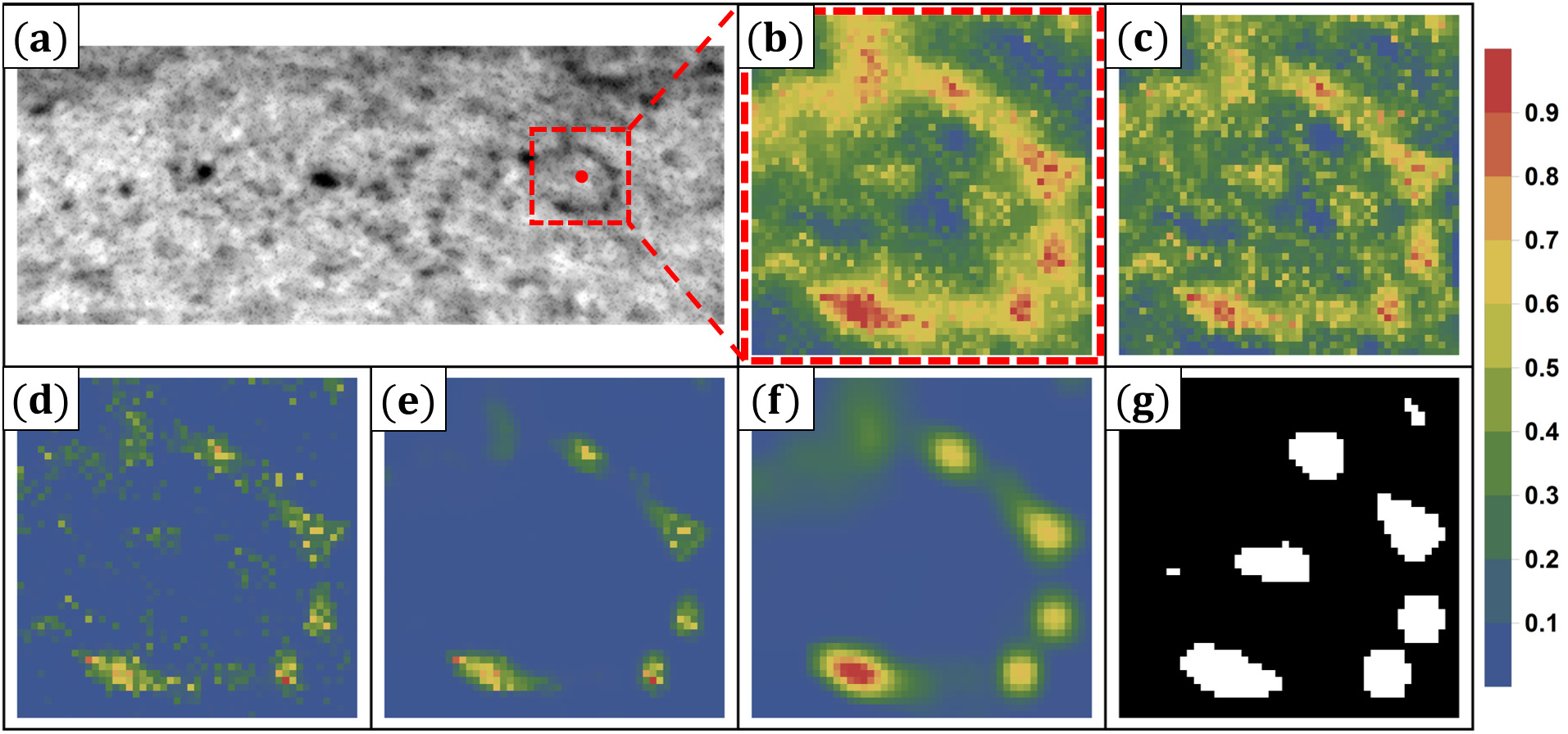}
			\caption{Local filtering applied to interrogation windows (IWs): (a) original image of the region of interest (Figure \ref{fig:fov-stdv-min-projections}) with a highlighted IW (red frame), (b) colorized luminance-inverted IW image; IW after sequentially applying (c) non-local means masking (NMM), (d) soft color tone map masking (SCTMM), (e) non-local means (NM) filtering, (f) mean filtering and (g) SCTMM followed by local adaptive binarization.}
			\label{fig:local-segmentation}
	\end{figure}

The stages of local filtering are shown in Figure \ref{fig:local-segmentation} and the filter structure is outlined in Algorithm \ref{alg:local-filter}. The luminance maps are inverted to make particles stand out as higher luminance regions, since by default, due to intense neutron flux absorption by particles, they appear in images as lower luminance zones. This results in an IW image as shown in Figure \ref{fig:local-segmentation}b. Next, non-local means masking (NMM) is performed (Figure \ref{fig:local-segmentation}c) to increase the contrast-to-noise ratio for particles and remove the "haze" (correlated noise due to unsharpness described in Section \ref{sec:experiment}), which is especially important for tightly-packed particle clusters. NMM transforms the original image $x$ into output $y$ as follows:

\begin{equation}
y = 2*x - w_\text{nm} * \text{NM}(x,r_\text{l},r_\text{p})
\label{eq:nm-correction}
\end{equation}

where $\text{NM}(x,r_\text{l},r_\text{p})$ is the non-local means (NM) filter \cite{non-local-means-filter}, $w_\text{nm}$ is the NM mask weight, and $r_\text{l}$ and $r_\text{p}$ are the filtering neighborhood and neighborhood comparison radii, respectively. This is similar in principle to unsharp masking, but utilizes the NM filter instead of the Gaussian filter. We observe that here NMM distinctly outperforms simple unsharp masking since the NM filter captures the correlated noise much better. Here we set $w_\text{nm} = 1.5$, $r_\text{l} = 1$ (pixels), $r_\text{p} = 5$. The best result is achieved when a noise power factor $p_\text{n} = 0.5$ is specified as well (estimated from the normalized luminance values for particles and "haze" (Figure \ref{fig:local-segmentation}b)). The utilized NM filter computes the normalized neighborhood weights $\tilde{w}$ for averaging as in \cite{non-local-means-filter-new-weight-function}:

\begin{equation}
    w_{ij} =
    \exp{ \left[
    - \text{max} \left( 0, ~
    \frac{1}{k_\sigma^2} \cdot
    \left(
    \frac{E_{ij}^2}{p_\text{n}} - 2  
    \right)
    \right)
    \right]
    }; ~~ \tilde{w}_{ij} = \frac{w_{ij}}{\text{max} (w_{ij})}; ~~ \tilde{w}_{00} = 1
\end{equation}

where $i$ and $j$ are the neighborhood indices, $E_{ij}$ is the Euclidean distance between neighborhoods and $k_\sigma$ is the filtering parameter. Here $k=0.75$.

Next, soft color tone map masking (SCTMM) was applied for background reduction (Figure \ref{fig:local-segmentation}d). SCTMM works by transforming an original normalized image $x$ to output $y$ in the following way:

\begin{equation}
y = x * 
\left( 
x - \left( 1 - \text{CTM} (x,c) \right) 
\right)
\label{eq:ctm-correction}
\end{equation}

where $\text{CTM} (x,c)$ is the color tone mapping operation and $c$ is the luminance compression factor. The motivation and principles behind SCTMM are explained in detail in \cite{birjukovs2021resolving}. Here we set $c = 0.65$. Afterwards, NM filtering is performed (Figure \ref{fig:local-segmentation}e) with $r_\text{l} = 2$ and $r_\text{p} = 10$ ($p_\text{n}$ automatically derived from the neighborhood squared standard deviation of luminance), followed by the mean filter (Figure \ref{fig:local-segmentation}f) with a 2-pixel radius. Note that throughout the IW filtering procedure, images are re-normalized after each filtering stage.

\begin{figure}[h]
			\centering
			\includegraphics[width=1\linewidth]{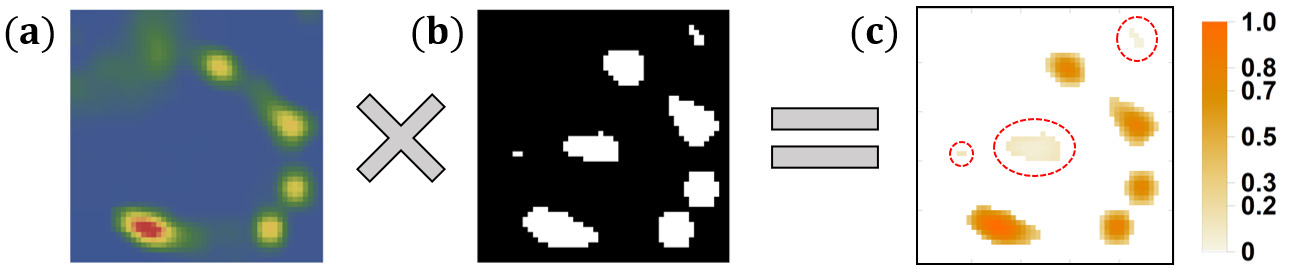}
			\caption{An illustration of intensity-based false positive filtering: (a) local filter output for an IW (Figure \ref{fig:local-segmentation}f) is multiplied by (b) the IW particle mask (Figure \ref{fig:local-segmentation}g) and $\left< I \right> \cdot \max (I)$ is computed for (c) the masked luminance $I$ map of each particle, as in \cite{birjukovs2021resolving}. The particles highlighted with red dashed lines were identified as false positives.}
			\label{fig:false-positive-filtering}
	\end{figure}

Filtered images are then subjected to the segmentation procedure outlined in Algorithm \ref{alg:local-segmentation}. Here local adaptive (LA) binarization (mean- and deviation-based) \cite{local-adaptive-thresholding} is used because global thresholding yields very unstable particle detection in filtered IWs due to their dissimilar SNR and CNR, and thus post-filtering luminance distributions. LA binarization, however, is susceptible to the edges of low-luminance particles many of which are potential false positives. For this reason, a special luminance-based false positive filtering procedure was used as in \cite{birjukovs2021resolving} with minor modifications. The underlying operations are stated in Algorithm \ref{alg:false-positive-luminance-filter} and its application is illustrated in Figure \ref{fig:false-positive-filtering}.

\begin{algorithm}

    \nonl \textbf{Input:} A normalized filtered IW image (Algorithm \ref{alg:local-filter} and Figure \ref{fig:local-segmentation}f):
    
    Apply SCTMM
    
    Local adaptive (LA) binarization
    
    Remove border components

    \nonl \textbf{Output:} Particle segment mask for the IW (Figure \ref{fig:local-segmentation}g)
    
\caption{Local (IW) segmentation}
\label{alg:local-segmentation}
\end{algorithm}

\begin{algorithm}

    \nonl \textbf{Input:} \\
    \begin{itemize}[noitemsep,topsep=0pt]
    \popline  \item Filtered IWs (Algorithm \ref{alg:local-filter})
        \item IW particle masks (Algorithm \ref{alg:local-segmentation})
    \end{itemize}
    
    \nl Multiply filtered IWs by the corresponding particle binary masks
    
    \nl Normalize the images
    
    \nl Compute $\left< I \right> \cdot \max (I)$ for all particles in IWs
    
    \nl \underline{\textit{Thresholding for all particles:}}
    
    \eIf{$\left< I \right> \cdot \max (I) < \eta; ~~ \eta \in [0;1]$ (user-defined)}{
    Flag the particle as a false positive
    }{
    Nothing
    }
    
    \nl Remove the identified false positives from the particle detection masks
    
    \textbf{Output:} IW particle detection masks without the detected false positives
    
\caption{Luminance map-based false positive filtering for IW particle detection masks}
\label{alg:false-positive-luminance-filter}
\end{algorithm}

The motivation and principles behind Algorithm \ref{alg:false-positive-luminance-filter} are provided in \cite{birjukovs2021resolving}. Note that here we set the LA binarization neighborhood radius to 10 (values above the mean within the radius are set to 1, while the rest are assigned 0), $c = 0.65$ for SCTMM, and $\eta = 0.1$ is used for false positive filtering. Border components are removed to avoid artifacts and artificial particle splitting.

\begin{figure}[h]
			\centering
			\includegraphics[width=1\linewidth]{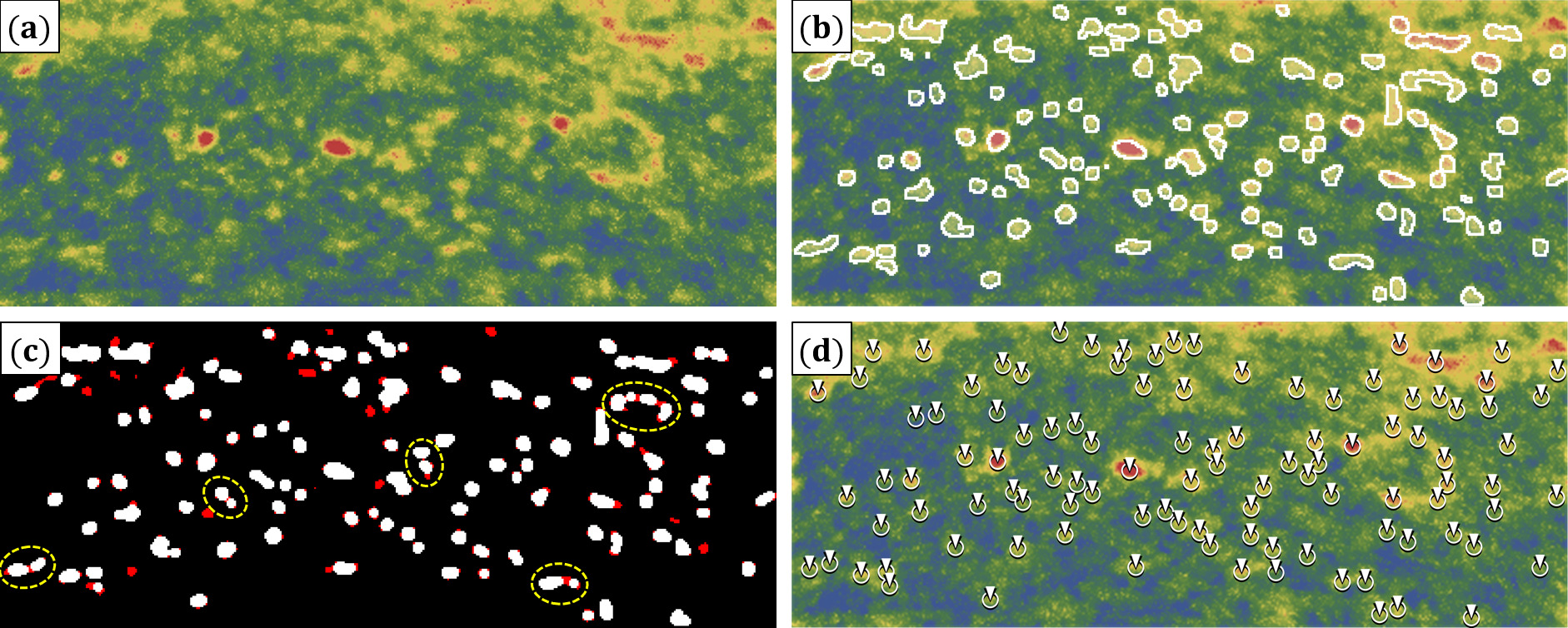}
			\caption{(a) Colorized luminance-inverted original image, (b) assembled global mask with detected particles overlaid on top of (a), (c) particle segments after area thresholding and small-radius morphological opening using disk elements and (d) particle centroids from (c) indicated with white arrows overlaid on top of (a). Note the red-colored segments in (c) -- these are the elements removed from (b) by area thresholding and erosion; yellow dashed lines in (c) indicate the segments that were resolved into fragments.}
			\label{fig:resolving-particles}
	\end{figure}

Once image filtering, segmentation and luminance-based false positive filtering are complete for the IWs from the original images, the filtered IW particle masks must be assembled into full FOV masks (Algorithm \ref{alg:image-proc-framework}). Figure \ref{fig:resolving-particles} shows the stages of this process. IW particle masks for every FOV image are mapped into the FOV (black background) and summed (Figure \ref{fig:resolving-particles}b). Then segment area thresholding and morphological opening (disk structural elements) \cite{images-mathematical-morphology} are performed (Figure \ref{fig:resolving-particles}c). Here the minimum area threshold is set to 5 pixels and the opening radius was set to 2 pixels. Finally, particle centroids (Figure \ref{fig:resolving-particles}d) are computed for the remaining particle segments (4-connectivity is used).

It is important to note that persistent artifacts within images may be a problem in that they might introduce systematic errors into trajectories output by a tracing algorithm. Notice that one such artifact is present in Figure \ref{fig:fov-stdv-min-projections}a -- a black spot in the right part of the FOV, which is a particle stuck to the flow channel. In such cases, we find that removing these artifacts with texture synthesis-based inpainting \cite{wolfram-mathematica-inpaint} is an effective solution. The artifacts in the considered images are readily segmentable from the mean projection over time for a sequence of images using Otsu binarization \cite{otsu-thresholding}.

\subsection{Parallelization \& performance}

The image processing pipeline as outlined above was implemented in \textit{Wolfram Mathematica} using its parallel computing functionality. Due to a large number of IWs per image ($\sim 10^2$) and the small relative area of an IW ($\sim 3.8 \%$), it was decided to parallelize IW processing for individual images, thus performing sets of parallel computations, as many as there are images in a sequence. IW generation is performed in serial mode, while local filtering and particle segmentation (Algorithm \ref{alg:image-proc-framework}, Steps 4 and 5) are parallelized. To speed up false positive filtering (Algorithm \ref{alg:false-positive-luminance-filter}), it is split into a sequence of steps, each individually parallelized, as outlined in Algorithm \ref{alg:parallelize-false positive-filtering}. Afterwards, the assembly of the full FOV masks is performed in parallel (image composition and addition), while area thresholding and morphological opening are performed in serial mode.

\begin{algorithm}

    Generate masked luminance maps for all IWs -- multiplication of filtered images and particle segment masks
    
    Compute $\left< I \right> \cdot \max (I)$ for all resulting segment intensity maps
    
    Compare the output against $\eta$ and flag false positives for all IWs
    
    Get segment masks for all IWs
    
    Map the particle masks of the true positives to all respective IWs
    
\caption{Splitting Algorithm \ref{alg:false-positive-luminance-filter} into sequential parallelized stages}
\label{alg:parallelize-false positive-filtering}
\end{algorithm}

The implementation as presented in this paper was tested on two machines (\textit{Windows 10}):

\begin{itemize}[noitemsep,topsep=0pt]
    \item \href{https://ark.intel.com/content/www/us/en/ark/products/198017/intel-core-i9-10980xe-extreme-edition-processor-24-75m-cache-3-00-ghz.html}{Intel Core i9-10980XE} (18 cores/36 threads) with 256 Gb $2933$ MHz DDR4 RAM
    \item \href{https://ark.intel.com/content/www/us/en/ark/products/199335/intel-core-i7-10700k-processor-16m-cache-up-to-5-10-ghz.html}{Intel Core i7-10700K} (8 cores/16 threads) with 32 Gb 3200 MHz DDR4 RAM
\end{itemize}

The code was first tested using a sequence of 1500 FOV images. Three test cases were run multiple times each: using all available parallel threads (hyperthreading was used since all the underlying operations for IWs are independent) on the \textit{Core i9} and \textit{Core i7} systems, and using the \textit{Core i9} CPU at $50\%$ parallel processing capacity. The results are summarized in Table \ref{tab:code-performance-tests}.

\begin{table}[!h]
\label{tab:code-performance-tests}
\begin{center}
\begin{tabular}{| c | c | c | c | c | c | c |}
\hline
\textbf{System} & \textbf{Threads} & \textbf{RAM used} & \textbf{Wall time ($\tau$)} & \textbf{Algorithms \ref{alg:local-filter} \& \ref{alg:local-segmentation}} & \textbf{Algorithm \ref{alg:false-positive-luminance-filter}} & \textbf{Mask assembly} \\ 
\hline
Core i9 & 36 & $<70$ Gb & 1.26 hrs & $59\%~ \tau$ & $28\%~ \tau$ & $13\%~ \tau$\\ \hline
Core i7 & 16 & $<32$ Gb & 1.80 hrs & $80\%~ \tau$ & $13.5\%~ \tau$ & $6.5\%~ \tau$\\ \hline
Core i9 & 18 & $<40$ Gb & 1.69 hrs & $72\%~ \tau$ & $19\%~ \tau$ & $9.0\%~ \tau$\\ \hline
\end{tabular}
\end{center}
\caption{The results of the image processing code benchmarks. RAM utilization accounts for the  system processes.}
\end{table}

In both cases with all available parallel threads used,  the CPU utilization for Algorithms \ref{alg:local-filter} and \ref{alg:local-segmentation} was consistently at $100\%$. For the \textit{Core i9} system with all threads utilized, all stages of Algorithm \ref{alg:parallelize-false positive-filtering} combined exhibit mean CPU utilization of $\sim 39\%$ on average, with $\sim 28\%$ at minimum and $\sim 83\%$ at maximum. The global mask assembly runs with $\sim 81\%$ CPU utilization on average. CPU utilization for the \textit{Core i7} system was greater for both Algorithm \ref{alg:parallelize-false positive-filtering} and global mask assembly: $\sim 59\%$ and $\sim 90\%$ on average, respectively. The discrepancies in processing time proportions and CPU utilization between the \textit{Core i9} and \textit{Core i7} systems are largely due to a considerable difference in the number of cores in favor of the \textit{Core i9} machine and the better single-core performance of the \textit{Core i7} machine. The speedup factor between the two systems systems running all available threads is $\sim 1.43$.

\subsection{Particle detection density}

\begin{figure}[h]
			\centering
			\includegraphics[width=0.625\linewidth]{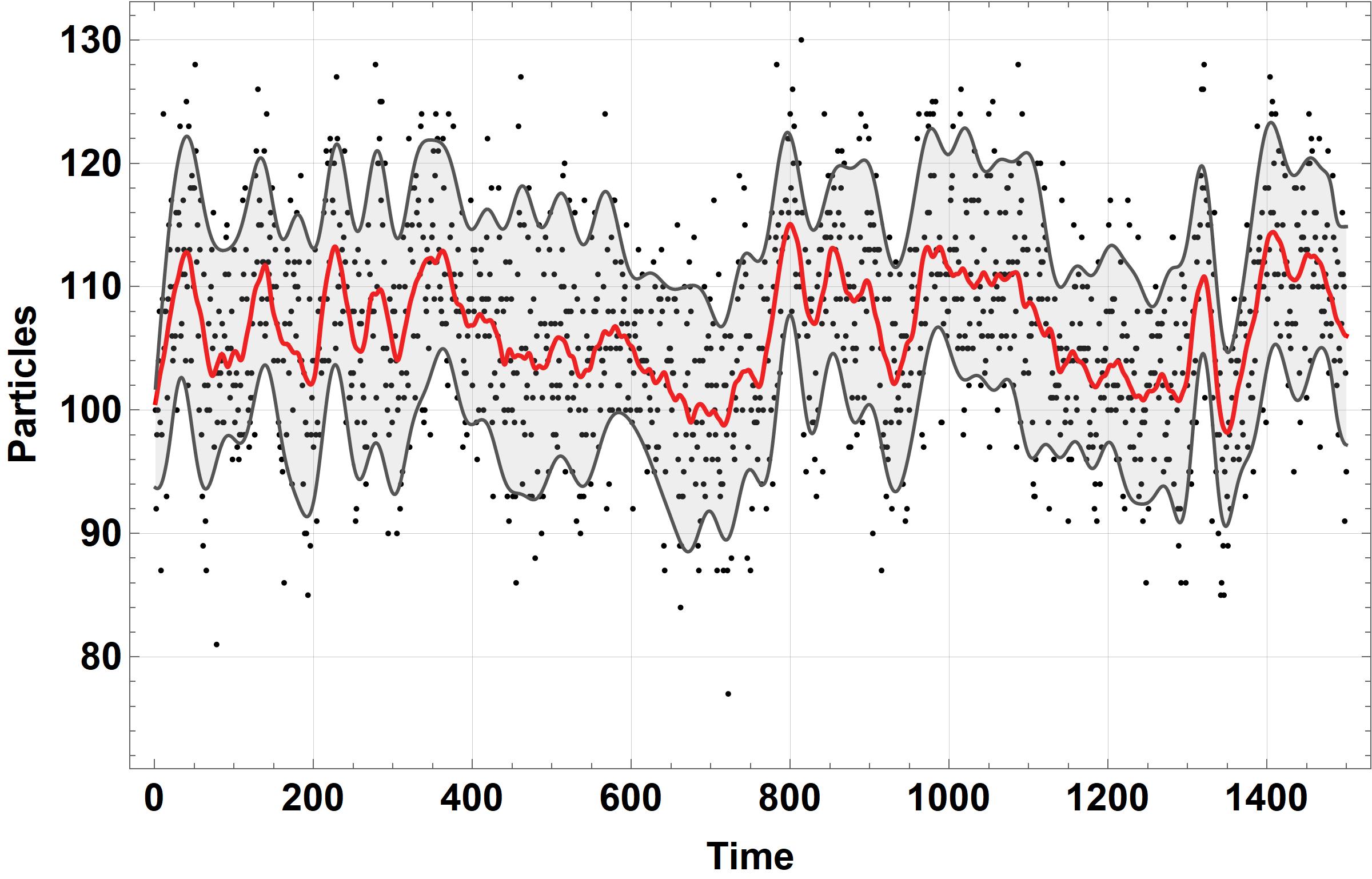}
			\caption{Particle count per frame (100 FPS, black) over a 1500-frame image sequence. The red curve is the averaged trend obtained via Gaussian total variation (TV) filtering (regularization parameter equal to 2) \cite{total-variation-rof-model} and the statistically significant ($q=0.9$ quantile) value ranges about the averaged curve are indicated with the light-gray envelope. The envelope is derived by filtering the \href{https://github.com/antononcube/MathematicaForPrediction/blob/master/QuantileRegression.m}{quantile spline envelopes} \cite{antonov-qse} for data with the same TV filter as the data.}
			\label{fig:particle-count}
	\end{figure}

\begin{figure}[h]
			\centering
			\includegraphics[width=0.775\linewidth]{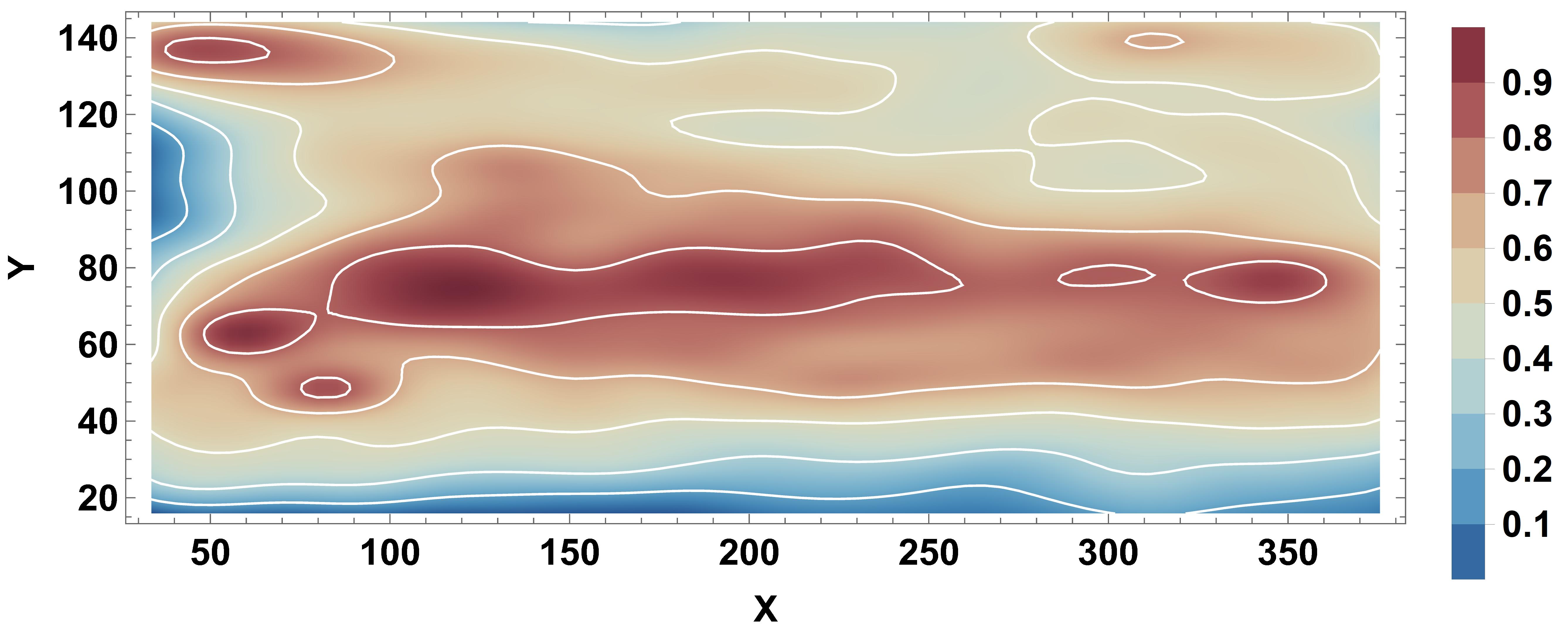}
			\caption{Normalized area density of particle detection events within the FOV (Figure \ref{fig:fov-stdv-min-projections}) over a 1500-frame image sequence. Density isolines are shown in white. The density map was computed using a Gaussian kernel over the count area density with Silverman's bandwidth estimation.}
			\label{fig:detection-density-all-frames}
	\end{figure}

Figures \ref{fig:particle-count} and \ref{fig:detection-density-all-frames} show the particle detection density over frames (equivalently, in time with 100 FPS) and space for a 1500-frame image sequence. Particle count per frame (Figure \ref{fig:particle-count}) is on average $\sim 105$ with a $\sim 10\%$ deviation, indicating consistency in particle detection.

Figure \ref{fig:detection-density-all-frames} indicates that particle detection density is considerably greater within the wake of the cylindrical obstacle -- this makes sense intuitively, since particles entrapped in or travelling through the wake flow zone are slower and have longer residence times within the FOV than the particles travelling with the mean flow about the obstacle, to the top and bottom of the FOV. Hence, more detection events per unit area are generated in the wake flow region of the FOV. This means that generated detection events are physically consistent with what one would expect from the studied system.

\section{Particle tracking}

Once particle centroids were obtained for all images, tracing was carried out with the MHT-X algorithm that we have previously developed \cite{zvejnieks2021mhtx}. Here we present extensions to MHT-X that are utilized for tracing of dense particle flow in the conditions as seen in the present experiment.

MHT-X requires the definitions of an extrapolation method, \textit{association conditions}, \textit{association constraints} and \textit{statistical functions}. Since in this case splitting and merging of particles does not occur in the experiment, only particle translation, entry and exit events must be considered. This makes the association constraints redundant and the split/merge statistical functions obsolete.

The association condition is a logical expression that determines whether two trajectories can be associated. The conditions from the original paper are used with the following changes. First, the association constraints on linear acceleration and deflection from \cite{zvejnieks2021mhtx}  are used, since constraints on motion are still desirable. Second, the sphere of influence (SOI) approach used in \cite{zvejnieks2021mhtx}  to restrict association range (only objects with overlapping SOI can be associated) is modified. Instead of defining the SOI about particle locations in frames, a prediction model pinpoints the location $\vec{r}$ of the region that the particle is expected to move to within time $\Delta t$ and defines the SOI about that point. The prediction consists of the spline extrapolation for particle velocity $\vec{v}_s$ and that derived from projecting the particle image velocimetry (PIV) field computed in \cite{lappan2020a} onto particle centroids $\vec{v}_\text{piv}$:

\begin{equation}
    \vec r(t_0 \pm\Delta t) = \vec r(t_0) \pm \left( \alpha \cdot \vec{v}_s + (1-\alpha)\cdot\vec{v}_{\text{piv}} \right) \cdot \Delta t
    \label{eq:position-prediction}
\end{equation}
where $\alpha$ determines the prediction component weights.

The SOI radius $R$ is based on the velocity magnitude, with higher velocity magnitudes yielding a smaller SOI:

\begin{equation}
    R=R_\text{max}\cdot \exp \left(-\frac{1}{\lambda} \cdot ||\alpha \cdot \vec{v}_s + (1-\alpha) \cdot \vec{v}_\text{piv}||\right)
    \label{eq:soi-adaptive-radius}
\end{equation}
where $R_\text{max}$ is the upper limit for the SOI radius and $\lambda$ is a control parameter. If $\vec{v}_s$ is undefined, it and $\alpha$ are set to 0.

This effectively assumes that particles with higher velocities are more difficult to deflect and vice versa, emulating cones of vision for moving particles. If two such cones overlap, an association is formed.

Exit and entry event statistical functions are kept as in \cite{zvejnieks2021mhtx}, except the horizontal x-axis is now the primary one. A model closely resembling the association condition has been implemented for translational motion associations. The translation likelihood estimator consists of three components determined by the location, the linear acceleration and the change in the motion direction.

The location-based likelihood compares the predicted location to the hypothesized location:
\begin{equation}
    p_\text{pos}=\mathcal{N}(\delta r, 0, \sigma_\text{pos}\cdot\Delta t)
\end{equation}
where $\mathcal{N} (x, \mu, \sigma)$  is a normalized Gaussian distribution with its mean $\mu$ and standard deviation $\sigma$; $\delta r$ is the absolute difference between positions due to the prediction and the hypothesis. The acceleration-based likelihood is calculated as follows:
\begin{equation}
    p_\text{acc}=\mathcal{N}(a, 0, \sigma_{a})
\end{equation}
The direction-based likelihood component is designed to penalize large changes in the motion direction. The penalty scales with velocity magnitude:
\begin{equation}
    p_\text{dir}=\mathcal{N} \left(\delta \varphi, 0 , \pi\cdot \exp \left( {\frac{||\vec v||}{\lambda}} \right) \right)
    \label{eq:direction-prediction}
\end{equation}
where $\delta \varphi$ is the change in direction, $\vec{v}$ is velocity and $\lambda$ is a control parameter.

The overall likelihood is computed as a weighted sum of the above contributions.
\begin{equation}
    p=\beta_1 \cdot p_\text{pos} + (1-\beta_1)\cdot(\beta_2 \cdot p_\text{acc} + (1 - \beta_2 )\cdot p_\text{dir})
\end{equation}
where $\beta_1$ and $\beta_2$ are weights.

\section{Preliminary results}

\begin{figure}[h]
			\centering
			\includegraphics[width=1\linewidth]{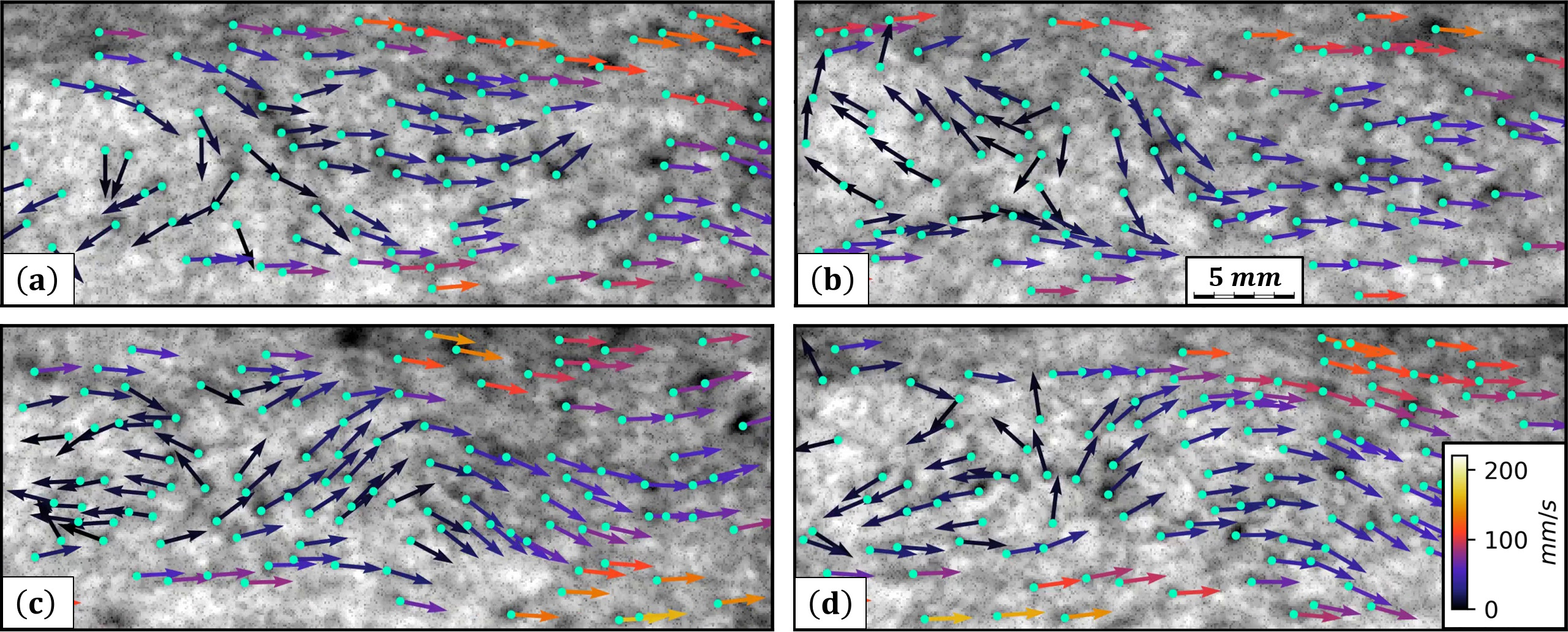}
			\caption{PIV field projected onto particle centroids at different time stamps. Note the scale bar in (b) and the velocity color bar (normalized for the entire image sequence) in (d).}
			\label{fig:piv-projection}
	\end{figure}

Before proceeding with tracking, the PIV velocity field $\vec{v}_\text{piv}$ obtained in \cite{lappan2020a} was interpolated and projected onto the positions of particles detected in each frame. Delaunay triangulation is performed for $\vec{v}_\text{piv}$ point grid and cubic interpolation is used for particle centroids that are within triangles formed by nearby $\vec{v}_\text{piv}$ grid points, while nearest neighbor interpolation is used otherwise (\href{https://docs.scipy.org/doc/scipy/reference/generated/scipy.interpolate.griddata.html}{\textit{SciPy}}). Interpolation is performed independently for both velocity components. Particle flow images with $\vec{v}_\text{piv}$ projections for particles are shown in Figure \ref{fig:piv-projection}. Note that, according to the $\vec{v}_\text{piv}$ field, many particles within the wake flow zone often travel in directions opposite (and sometimes normal, as seen in Figure \ref{fig:piv-projection}d) to the general flow direction. The obtained $\vec{v}_\text{piv}$ for particles is used in (\ref{eq:position-prediction}) and (\ref{eq:soi-adaptive-radius}) for motion prediction.

\begin{figure}[h]
			\centering
			\includegraphics[width=1\linewidth]{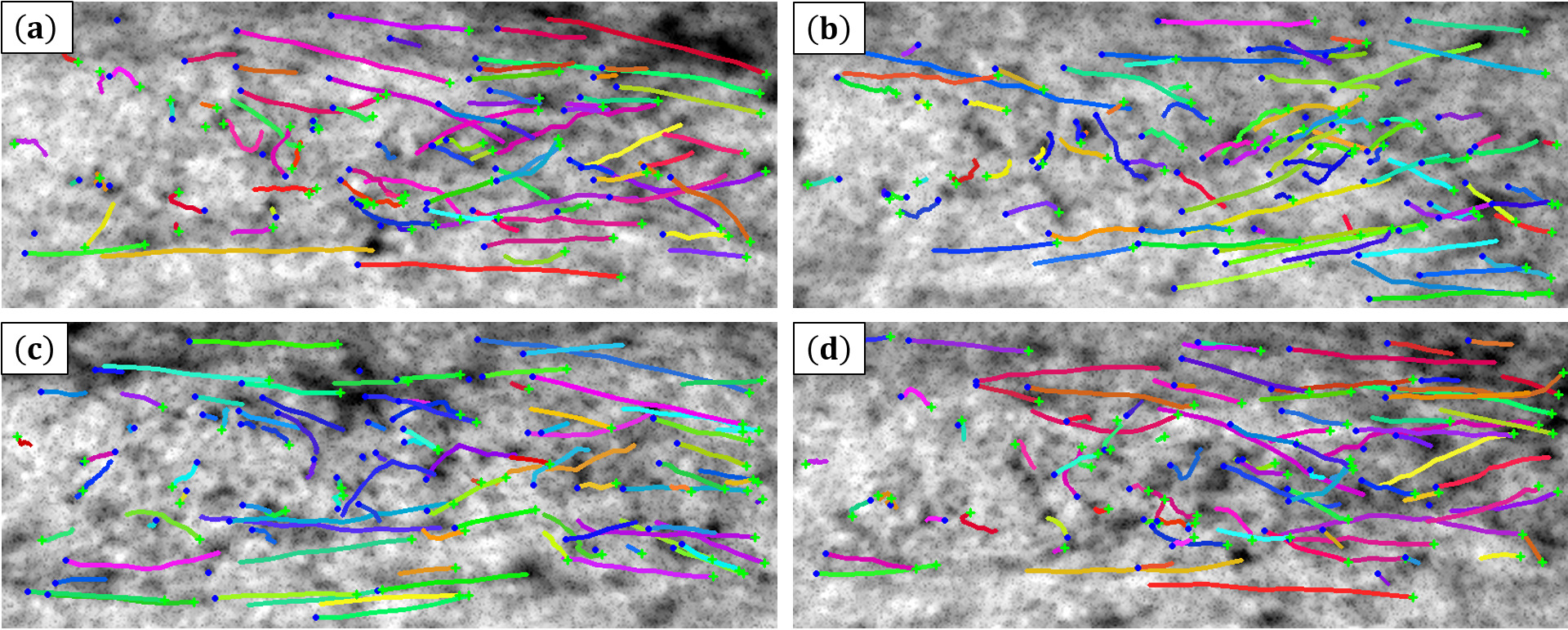}
			\caption{Snapshots of constructed particle trajectories (indicated with different colors) at different time stamps. Particle \textit{Entry} nodes are indicated with blue dots, while \textit{Exit} nodes are shown as green crosses. Each of the snapshots shows the last 15 segments of reconstructed trajectories. The scale is identical to that shown in Figure \ref{fig:piv-projection}.}
			\label{fig:particle-id-tracks}
	\end{figure}

\begin{figure}[h]
			\centering
			\includegraphics[width=1\linewidth]{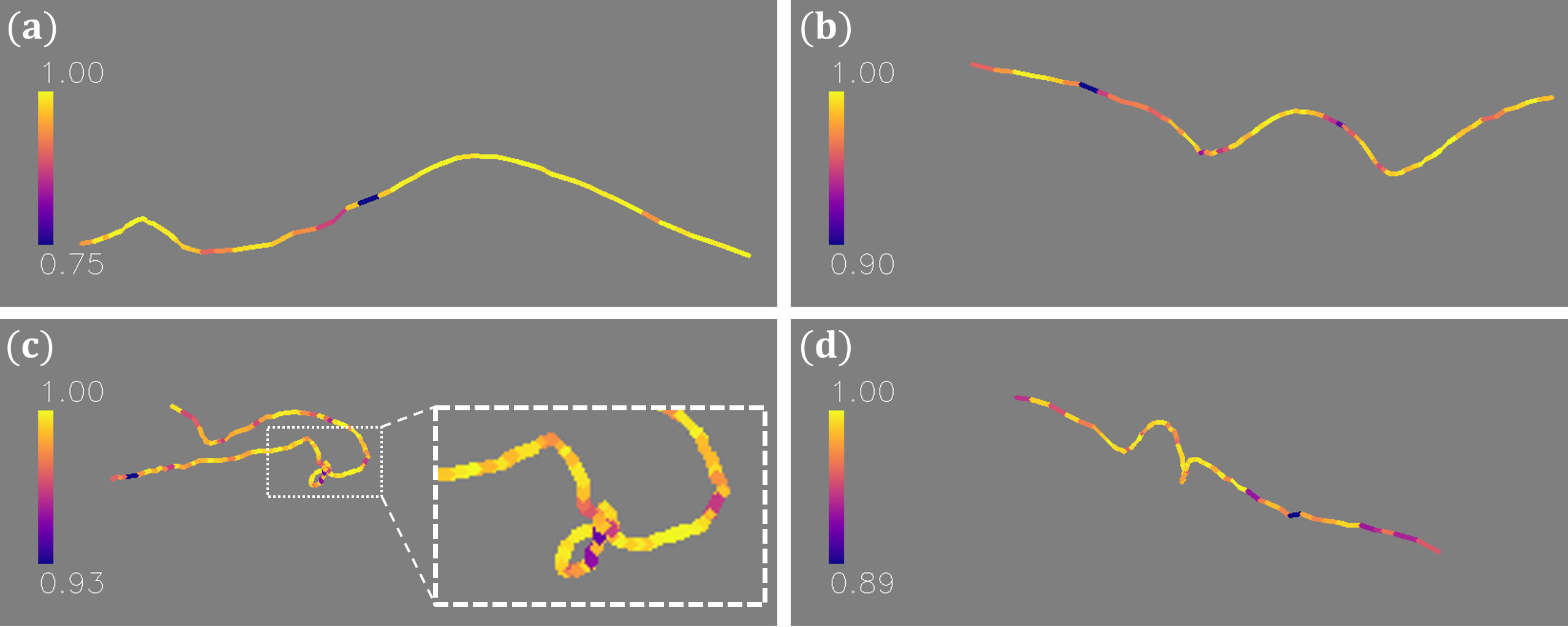}
			\caption{Representative particle trajectories reconstructed by MHT-X. Trajectory segments are color-coded by their likelihoods. The length scale is identical to that shown in Figure \ref{fig:piv-projection}.}
			\label{fig:particle-trajectories}
	\end{figure}

Consider Figures \ref{fig:particle-id-tracks} and \ref{fig:particle-trajectories} showing the results of applying MHT-X to the output of image processing. Figure \ref{fig:particle-id-tracks} shows some of the reconstructed trajectories within the FOV at four different time stamps. Note that only the last 15 segments of the constructed trajectories are shown. This limitation was introduced for visual clarity, but the trade-off is that the trajectories of slower particles in the wake flow zone are more difficult to show. Despite this, several things can be observed. First, note that trajectories are not broken near the right boundary of the FOV where an image artifact used to be before being removed via texture synthesis inpainting. Second, notice that even with the limitation on the number of segments visible at a time per trajectory, rather long particle tracks can be observed both within and outside the wake flow zone. Third, one can see, especially in Figures \ref{fig:particle-id-tracks}a and \ref{fig:particle-id-tracks}d, that densely packed trajectories that cross one another in close temporal proximity are correctly resolved. However, it is also evident that there are quite a few significantly fragmented trajectories, especially within the wake flow zone.

Figure \ref{fig:piv-projection} demonstrates the issue of particles being caught within the oscillating wake flow area exhibit both relatively small velocity magnitudes \textit{and} rapid changes in motion direction. This is problematic for the current MHT-X implementation since closely packed trajectories with low velocity magnitudes, according to (\ref{eq:soi-adaptive-radius}) and (\ref{eq:direction-prediction}), result in many feasible associations for trajectory connections. With the current spline-based trajectory extrapolation method \cite{zvejnieks2021mhtx} it is often the case that trajectory fragment mismatch is such that MHT-X opts to assign \textit{Exit} nodes to trajectories prematurely rather than reconstruct longer tracks from fragments. However, MHT-X is still able to resolve quite a number of physically meaningful and long trajectories, examples of which are shown in Figure \ref{fig:particle-trajectories}.

\begin{figure}[h]
			\centering
			\includegraphics[width=1\linewidth]{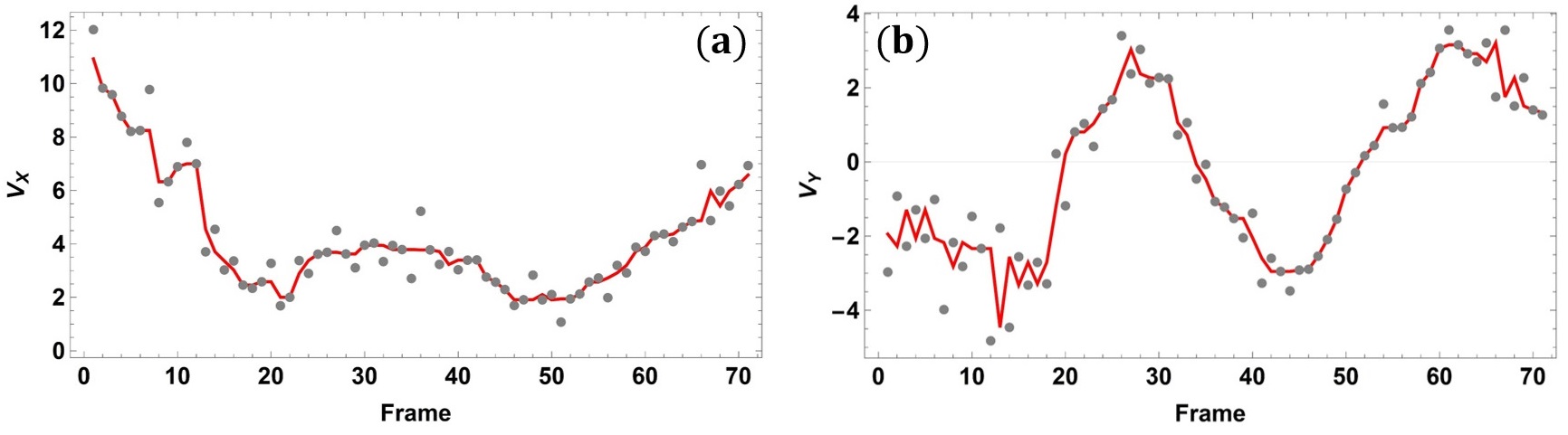}
			\caption{Velocimetry for the particle with the trajectory shown in Figure \ref{fig:particle-trajectories}b: velocity (a) $x$ and (b) $y$ components, pixels per frame. The gray dots are the MHT-X output and the red curves are the median-filtered (1-point radius) velocity components.}
			\label{fig:trajectory_velocity_components}
	\end{figure}

Figure \ref{fig:particle-trajectories} shows some of the longer trajectories recovered by MHT-X. In Figures \ref{fig:particle-trajectories}a and \ref{fig:particle-trajectories}b one can see trajectories of particles that passed by, interacted with and then departed from the wake flow zone. Notice in Figure \ref{fig:particle-trajectories}b that as the particle is briefly captured by the wake flow, its velocity becomes lower as indicated by significantly shorter trajectory segments seen in the middle of the FOV ($x$ direction) -- this is also seen in Figure \ref{fig:trajectory_velocity_components}.

Then, as the particle exits the wake flow, it is again accelerated by the channel flow. Figure \ref{fig:particle-trajectories}a, on the other hand, shows that the particle was not entrapped in the wake flow and traversed the FOV much more rapidly. The trajectory in Figure \ref{fig:particle-trajectories}c is the longest observed in terms of the number of segments. The underlying particle was first observed and entered the wake flow zone from the top of the FOV and then had a considerable residence time within the wake before the trajectory was broken off in the left part of the FOV. This particular trajectory is of note for several reasons: first, it clearly shows what is also seen in the PIV projection images -- particles captured by the wake behind the obstacle are often diverted towards the center of the wake flow zone and then their direction is reversed such that it is opposite to that of the mean channel flow \cite{sommer-4d-ptv}; second, note the fragment of this trajectory highlighted by a white dashed frame in Figure \ref{fig:particle-trajectories}c -- one can observe the particle trajectory forming a small loop. It is important that such motion with a low velocity in presence of other potentially interfering particles in the wake flow zone is nonetheless resolved with a high degree of confidence -- note the minimum segment likelihood is 0.93 (color bar to the left). Figure \ref{fig:particle-trajectories}d shows a similar trajectory except that its motion direction is not reversed during the residence time.

\begin{figure}[h]
			\centering
			\includegraphics[width=1\linewidth]{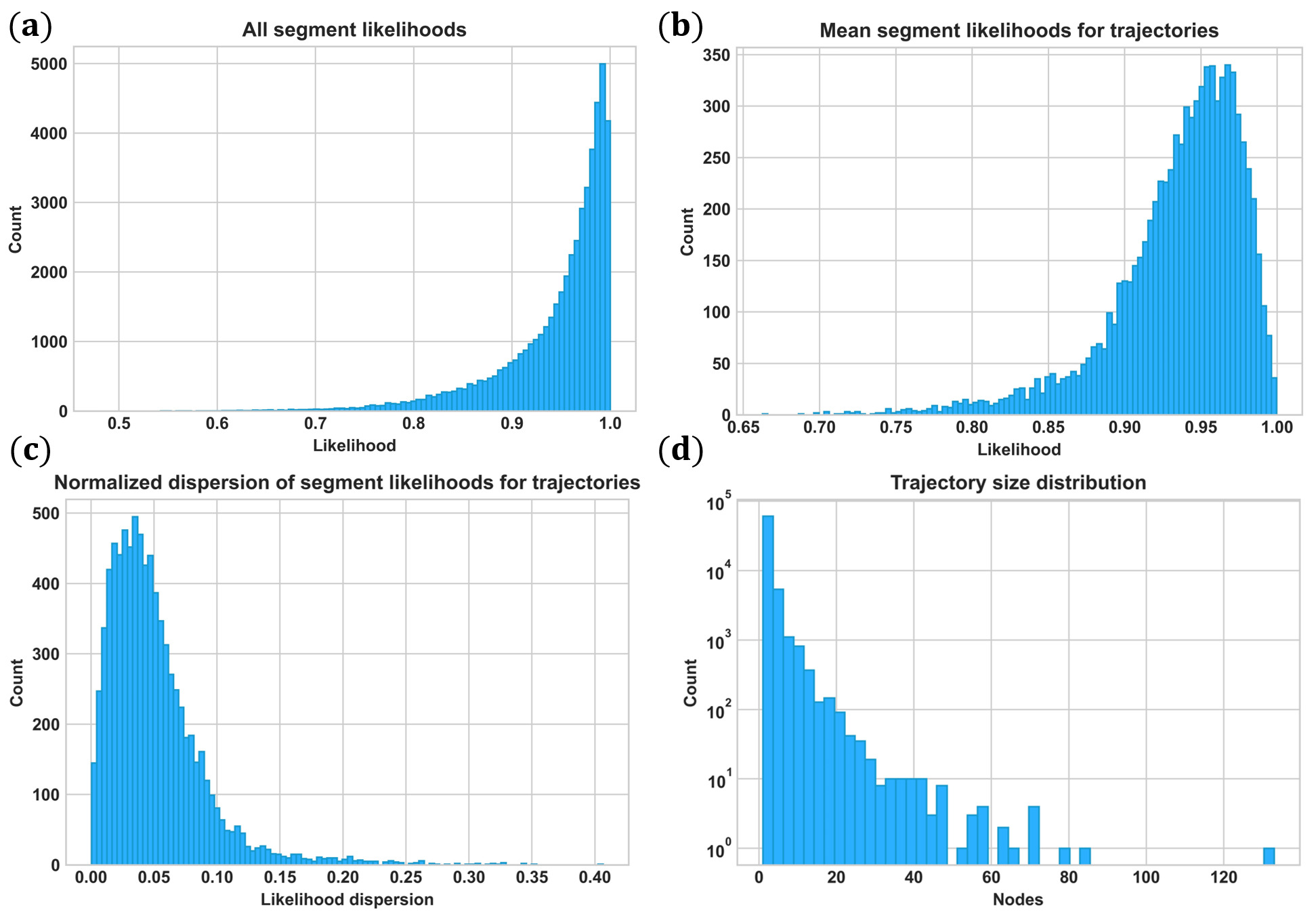}
			\caption{(a) The likelihoods of all trajectory segments, (b) mean segment likelihoods for all trajectories, (c) standard deviation of segment likelihoods within trajectories, normalized by mean likelihoods, and (d) node count for constructed trajectories. In (a-c) trajectories with 4+ nodes are considered.}
			\label{fig:particle-trajectory-stats}
	\end{figure}

To assess the quality of MHT-X output more quantitatively, several metrics were evaluated for recovered trajectories: segment likelihoods for all segments, mean segment likelihoods and normalized (with respect to mean) dispersion of likelihoods for trajectories, as well as the trajectory size distribution -- these are given in Figure \ref{fig:particle-trajectory-stats}. The first thing to note is that most of the segments for trajectories with 4+ nodes (trajectories with $<4$ nodes are not usable even for local PTV) have likelihoods mostly in excess of $0.9$ with a sharp maximum just below $1$ (Figure \ref{fig:particle-trajectory-stats}a). This is important since greater likelihoods generally imply lesser trajectory reconstruction ambiguity and greater confidence that the output is physically sound. Very high mean segment likelihoods for constructed trajectories (Figure \ref{fig:particle-trajectory-stats}b) and mean likelihood dispersion mostly within $10\%$ of the mean values (Figure \ref{fig:particle-trajectory-stats}c) also speak to the quality of the generated results. Finally, Figure \ref{fig:particle-trajectory-stats}d indicates that MHT-X produced a few hundred trajectories with $\sim 20$ nodes and tens of trajectories with 20+ nodes. Note that the trajectory with 130+ nodes seen to the right of the distribution is the one shown in Figure \ref{fig:particle-trajectories}c. While in-depth physical analysis of flow dynamics and how it affects particle trajectories -- their residence time within different regions of the FOV, trajectory curvature, etc. -- requires more tracks with 30+ nodes (more for slower particles captured by the wake) than are currently produced, shorter tracks can be used for local PTV which could potentially provide insights about the velocity field at smaller length scales than in the case of PIV.

\begin{figure}[h]
			\centering
			\includegraphics[width=0.65\linewidth]{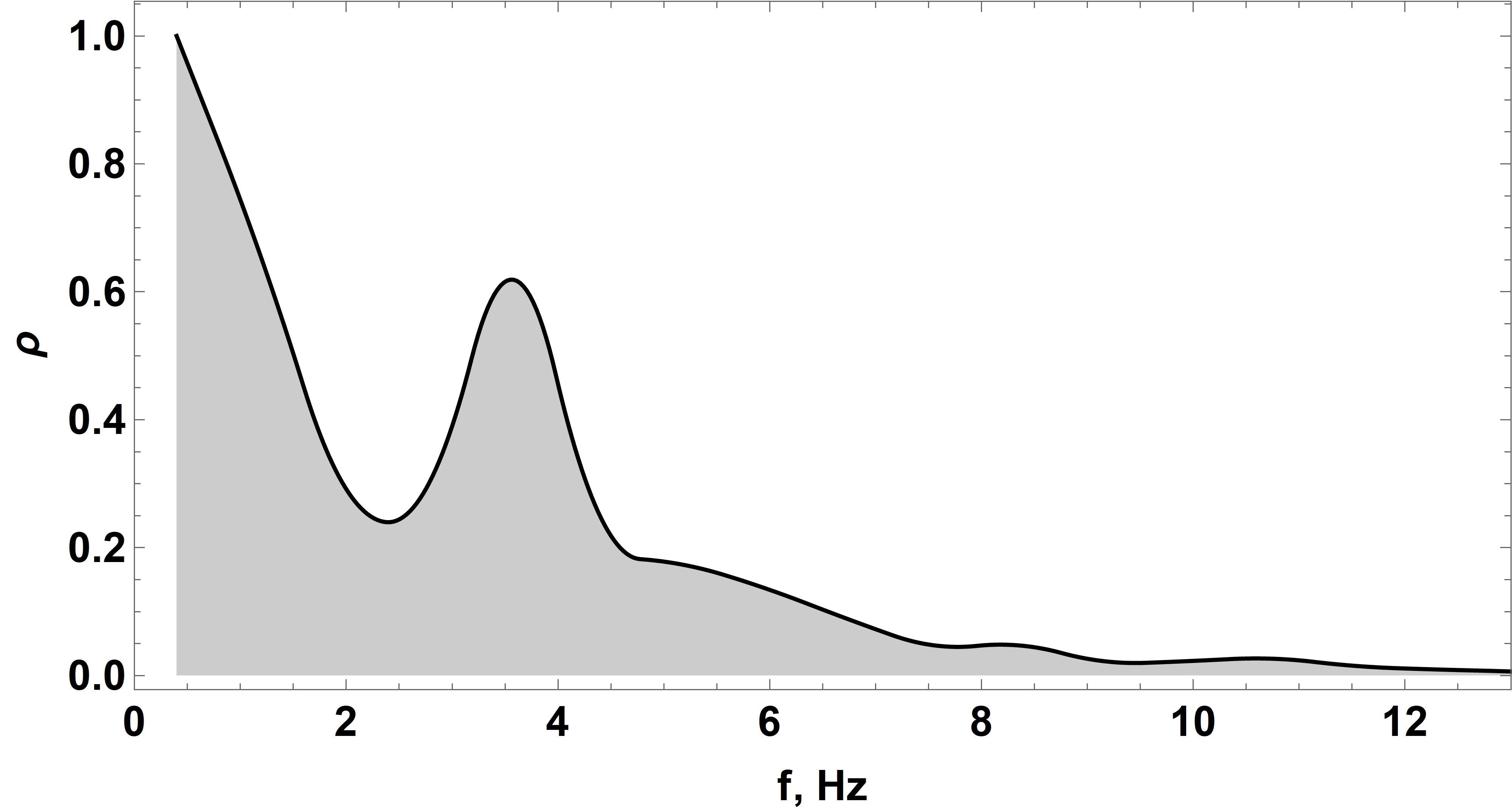}
			\caption{Smooth normalized density histogram (Scott binning, 2-nd order interpolation) of the first 10 dominant frequencies aggregated for 200 longest (by graph node count) trajectories.}
			\label{fig:trajectory-vel-X-freqency-spectrum}
	\end{figure}

Finally, it was previously assumed in \cite{lappan2020a} that the utilized gadolinium oxide particles are mostly passive tracers. However, it is worth testing this assumption by estimating the particle Stokes number ($Stk$) range. We first evaluate the Reynolds number for the cylindrical obstacle $Re_\text{c} = \rho_0 U_0 d_\text{c} / \mu_0$ with material properties as stated in Section \ref{sec:experiment}, and using the free-stream velocity from PIV $U_0 \in (10;14)~ cm/s$ \cite{lappan2020a}. This yields $Re_\text{c} \in (1466;2053)$, indicating a transitional or possibly turbulent cylinder wake flow regime -- this is also in line with the observed vortex shedding and wake flow oscillations. We then assess the particle $Re$ as $Re_\text{p} = \rho_0 |U_\text{rel}| d_\text{p} / \mu_0$ where $U_\text{rel}$ is the relative particle-flow velocity ($d_p$ range stated in Section \ref{sec:experiment}). Assuming that most of the particles have $|U_\text{rel}| \in (0;10)~ mm/s$, which is up to $\sim 33 d_\text{p}$ per second, one has $Re_\text{p} \lesssim 15$. Given the very coherent (even visually) motion seen in the cylinder wake and about the cylinder, it is very unlikely that $|U_\text{rel}| \sim 0.1 ~ m/s$ (stationary particle in the free-stream flow, $Re_\text{p} \sim 150$) can be expected except for rather rare instances. And even then, $Re_\text{p} = 150$ is well below the particle wake flow delaminarization threshold, while particles with $Re_\text{p} \lesssim 15$ should not even exhibit considerable flow separation \cite{clift-bubbles}. Therefore, the drag force on particles should not deviate too much from the Stokes law, which means one may set $Re_\text{p} \sim 1$ as an initial assumption. This implies that the particle time scale should be close to that of the wake flow, where the dominant time scale is associated with the vortex shedding frequency. Using $Re_\text{c}$ range, we estimate the Strouhal number ($Sr$) for the obstacle via $Sr = 0.198 \cdot (1 - 1.97/Re_\text{c})$ which results in $Sr \sim 0.1978$. $Stk = \tau_\text{p}/\tau_0$ which is the ratio of the particle and flow time scales, respectively. At the same time, $Sr = d_\text{c} / \tau_0 U_0$ and given $Re_\text{p} \sim 1$ one has $\tau_\text{p} = \rho_\text{p} d^2_\text{p} / 18 \mu_0$. This means that $Stk$ can be expressed through $Sr$ as $Stk = Sr \cdot \rho_\text{p} d^2_\text{p} U_0 / 18 \mu_0 d_\text{c}$, yielding $Stk \in (0.067; 0.262)$ and implying that the particles should be fairly good tracers under the above assumptions. Another approach to check this and also qualitatively validate the MHT-X performance is to consider the reconstructed trajectories, perform Fourier analysis of their velocity component dynamics, and check if the fundamental cylinder wake oscillation frequency $f_0$ corresponds to the frequency content extracted from trajectories $f_\text{t}$. $f_0$ can be estimated from $Sr$ as the inverse of $\tau_0$ -- here one has $f_0 \in (3.95; 5.54)~ \text{Hz}$. On the other hand, taking the first 10 dominant frequencies for 200 longest trajectories and assessing the probability density of the encountered frequencies reveals that there is a distinct peak with a full width at half maximum spanning $f_\text{t} \in (2.42; 4.23)~ \text{Hz}$ (Figure \ref{fig:trajectory-vel-X-freqency-spectrum}) which overlaps with $f_0$.

This indicates consistency between the reconstructed dynamics and the expected flow properties. Lastly, bear in mind that for $Re_\text{p}$ significantly greater than unity the Stokes drag law underestimates the drag force, meaning that the $Stk$ range estimate derived herein is likely overestimated.

\section{Conclusions \& outlook}

To summarize, we have developed and demonstrated an image processing methodology coupled with the modified MHT-X code for particle detection and tracking in images from dynamic neutron radiography of particle flow in a liquid metal channel. Preliminary results indicate that the proposed approach is feasible -- for local PTV as is, and for a more in-depth physical analysis of fully reconstructed trajectories, and thus wake flow, after the existing extrapolation scheme \cite{zvejnieks2021mhtx} is replaced with a better solution.

In this regard, we see several ways to improve/extend the functionality of MHT-X -- in addition to the points made in \cite{zvejnieks2021mhtx}, we propose the following:

\begin{itemize}[noitemsep,topsep=0pt]

    \item Use the Kalman filter for improved particle motion prediction by using three contributions: PIV- motion and PTV-predicted motion, as well as predictions derived by solving effective equations of motion for particles (Lagrangian approach).
    
    \item For higher-quality PIV-based predictions, it is planned to utilize divergence-free interpolation \cite{div-free-inter-and-motion-prediction, div-free-theory} to minimize errors stemming from the currently used polynomial interpolation ignoring the flow continuity constraint.
    
    \item A multi-level approach for divergence-free interpolation \cite{div-free-interpolation-multilevel} seems prospective for further improvement of PIV-based predictions.
    
    \item Better PTV-based predictions could also be generated applying divergence-free interpolation to sparse PTV velocity fields. However, since the PTV field is not defined on a regular grid like in the case of PIV, it would seem that the multi-level method is not readily applicable here and will have to be extended.
    
    \item Given the low particle Reynolds numbers seen in our case, one can use a linear drag model for effective equations of motion for particles. One could then solve these equations as proposed in \cite{div-free-inter-and-motion-prediction} using phase space volume contractivity preservation via splitting methods.
    
\end{itemize}

We expect that, with the outlined improvements implemented, the approach developed herein will enable us to generate many more longer trajectories. The latter can then be analyzed by computing residence time maps for the FOV, examining how trajectory curvature profiles over arc lengths are correlated with local flow dynamics, etc. It is likely that applying dynamic mode decomposition \cite{klevs2021dynamic} to PIV and PTV velocity fields could assist physical interpretation.

Finally, this study clearly demonstrates that flow analysis can be performed under conditions similar to what is shown in this paper -- as such, more complicated and realistic experiments of a similar nature can be conducted and successful extraction of physically meaningful information can be expected.

The image processing code is available at \textit{GitHub}: \href{https://github.com/Mihails-Birjukovs/Low_C-SNR_Particle_Detection}{Mihails-Birjukovs/Low\_C-SNR\_Particle\_Detection}. MHT-X can be found at \textit{GitHub} as well: \href{https://github.com/Peteris-Zvejnieks/MHT-X}{Peteris-Zvejnieks/MHT-X}.

\section*{Acknowledgments}

This research is a part of the ERDF project ”Development of numerical modelling approaches to study complex multiphysical interactions in electromagnetic liquid metal technologies” (No. 1.1.1.1/18/A/108) and is based on experiments performed at the Swiss spallation neutron source SINQ, Paul Scherrer Institute, Villigen, Switzerland. The authors acknowledge the support from Paul Scherrer Institut (PSI) and Helmholtz-Zentrum Dresden-Rossendorf (HZDR). The work is also supported by a DAAD Short-Term Grant (2021, 57552336) and the ANR-DFG project FLOTINC (ANR-15-CE08-0040, EC 217/3).

\printbibliography[title={References}]

@article{anderson1992,
  title = {The {{Ga}}-{{Sn}} (gallium-tin) system},
  author = {Anderson, T. J. and Ansara, I.},
  year = {1992},
  volume = {13},
  pages = {181--189},
  publisher = {{Springer Science and Business Media LLC}},
  doi = {10.1007/bf02667485},
  journal = {Journal of Phase Equilibria},
  number = {2}
}

@article{blau2009,
  title = {The {{Swiss}} spallation neutron source {{SINQ}} at {{Paul Scherrer Institut}}},
  author = {Blau, B. and Clausen, K. N. and Gvasaliya, S. and Janoschek, M. and Janssen, S. and Keller, L. and Roessli, B. and Schefer, J. and {Tregenna-Piggott}, P. and Wagner, W. and Zaharko, O.},
  year = {2009},
  volume = {20},
  pages = {5--8},
  doi = {10.1080/10448630903120387},
  journal = {Neutron News},
}

@article{div-free-interpolation-multilevel,
  title = {Multilevel interpolation of divergence-free vector fields},
  author = {Farrell, Patricio and Gillow, Kathryn and Wendland, Holger},
  year = {2017},
  volume = {37},
  pages = {332--353},
  doi = {10.1093/imanum/drw006},
  journal = {IMA Journal of Numerical Analysis},
  number = {1}
}

@article{kaestner2011,
  title = {The {{ICON}} beamline - {{A}} facility for cold neutron imaging at {{SINQ}}},
  author = {Kaestner, A. P. and Hartmann, S. and K{\"u}hne, G. and Frei, G. and Gr{\"u}nzweig, C. and Josic, L. and Schmid, F. and Lehmann, E. H.},
  year = {2011},
  volume = {659},
  pages = {387--393},
  doi = {10.1016/j.nima.2011.08.022},
  journal = {Nuclear Instruments and Methods in Physics Research Section A - Accelerators, Spectrometers, Detectors and Associated Equipment},
  number = {1}
}

@article{lappan2020a,
  title = {Neutron radiography of particle-laden liquid metal flow driven by an electromagnetic induction pump},
  shorttitle = {{{NR}}},
  author = {Lappan, Tobias and Sarma, Martins and Heitkam, Sascha and Trtik, Pavel and Mannes, David and Eckert, Kerstin and Eckert, Sven},
  year = {2020},
  volume = {56},
  pages = {167--176},
  doi = {10.22364/mhd.56.2-3.8},
  journal = {Magnetohydrodynamics},
  number = {2/3}
}

@book{lide2019,
  title = {{{CRC Handbook}} of {{Chemistry}} and {{Physics}}},
  editor = {Lide, David R},
  year = {2019},
  publisher = {{CRC Press}},
  address = {{Boca Raton}},
  isbn = {978-1-138-36729-6},
  language = {English}
}

@book{martienssen2005,
  title = {Springer {{Handbook}} of {{Condensed Matter}} and {{Materials Data}}},
  author = {Martienssen, W. and Warlimont, Hans},
  year = {2005},
  publisher = {{Springer}},
  address = {{Berlin, Heidelberg}},
  isbn = {978-3-540-44376-6}
}

@misc{PSI_LinAttCoeffTher,
  title = {Thermal neutron data \textendash{} {{Table}} of linear attenuation coefficients for thermal neutrons (25 {{meV}})},
  author = {{Paul Scherrer Institut (PSI)}},
  howpublished = {https://www.psi.ch/sites/default/files/import/niag/LinksEN/Thermal\_Attenu\_Coeff.pdf}
}

@article{sears1992,
  title = {Neutron scattering lengths and cross sections},
  author = {Sears, Varley F.},
  year = {1992},
  volume = {3},
  pages = {26--37},
  doi = {10.1080/10448639208218770},
  journal = {Neutron News},
  number = {3}
}

@article{div-free-inter-and-motion-prediction,
  title = {Computational geometric methods for preferential clustering of particle suspensions},
  author = {Tapley, Benjamin K. and Andersson, Helge I. and Celledoni, Elena and Owren, Brynjulf},
  year = {2021},
  month = feb,
  archiveprefix = {arXiv},
  eprint = {1907.11936},
  eprinttype = {arxiv}
}

@article{div-free-theory,
  title = {Divergence-free kernel methods for approximating the stokes problem},
  author = {Wendland, Holger},
  year = {2009},
  volume = {47},
  pages = {3158--3179},
  doi = {10.1137/080730299},
  journal = {SIAM Journal on Numerical Analysis},
  number = {4}
}

@misc{zvejnieks2021mhtx,
      title={MHT-X: Offline Multiple Hypothesis Tracking with Algorithm X}, 
      author={Peteris Zvejnieks and Mihails Birjukovs and Martins Klevs and Megumi Akashi and Sven Eckert and Andris Jakovics},
      year={2021},
      eprint={2101.05202},
      archivePrefix={arXiv},
      primaryClass={cs.CV}
}

@article{total-variation-rof-model,
title = {Nonlinear total variation based noise removal algorithms},
journal = {Physica D: Nonlinear Phenomena},
volume = {60},
number = {1},
pages = {259-268},
year = {1992},
issn = {0167-2789},
doi = {https://doi.org/10.1016/0167-2789(92)90242-F},
url = {https://www.sciencedirect.com/science/article/pii/016727899290242F},
author = {Leonid I. Rudin and Stanley Osher and Emad Fatemi}
}

@misc{antonov-qse,
      title={Quantile regression with B-splines}, 
      author={Anton Antonov},
      year={2014},
      url={https://mathematicaforprediction.wordpress.com/2014/01/01/quantile-regression-with-b-splines}
}

@article{birjukovsPhaseBoundaryDynamics2020,
	title = {Phase boundary dynamics of bubble flow in a thick liquid metal layer under an applied magnetic field},
	volume = {5},
	doi = {10.1103/PhysRevFluids.5.061601},
	journal = {Physical Review Fluids},
	shortjournal = {Physical Review Fluids},
	author = {Birjukovs, Mihails and Dzelme, Valters and Jakovics, Andris and Thomsen, Knud and Trtik, Pavel},
	date = {2020-06-18},
	year = {2020}
}

@article{birjukovsArgonBubbleFlow2020,
	title = {Argon bubble flow in liquid gallium in external magnetic field},
	volume = {63},
	doi = {10.3233/JAE-209116},
	pages = {1--7},
	journal = {International Journal of Applied Electromagnetics and Mechanics},
	shortjournal = {International Journal of Applied Electromagnetics and Mechanics},
	author = {Birjukovs, Mihails and Dzelme, Valters and Jakovics, Andris and Thomsen, Knud and Trtik, Pavel},
	date = {2020-04-29},
	year = {2020}
}

@article{baakeNeutronRadiographyVisualization2017,
	title = {Neutron radiography for visualization of liquid metal processes: bubbly flow for {CO}2 free production of Hydrogen and solidification processes in {EM} field},
	volume = {228},
	issn = {1757-899X},
	url = {https://doi.org/10.1088%2F1757-899x%2F228%2F1%2F012026},
	doi = {10.1088/1757-899X/228/1/012026},
	shorttitle = {Neutron radiography for visualization of liquid metal processes},
	pages = {012026},
	journal = {{IOP} Conference Series: Materials Science and Engineering},
	shortjournal = {{IOP} Conf. Ser.: Mater. Sci. Eng.},
	author = {Baake, E. and Fehling, T. and Musaeva, D. and Steinberg, T.},
	urldate = {2020-11-05},
	date = {2017-07},
	year = {2017},
	langid = {english},
	note = {Publisher: {IOP} Publishing}
}

@article{casting-euler-musig,
author = {Liu, Zhongqiu and Li, Linmin and Qi, Fengsheng and Li, Baokuan and Jiang, Mao Fa and Tsukihashi, Fumitaka},
year = {2015},
month = {02},
pages = {},
title = {Population Balance Modeling of Polydispersed Bubbly Flow in Continuous-Casting Using Multiple-Size-Group Approach},
volume = {46},
journal = {Metallurgical and Materials Transactions B},
doi = {10.1007/s11663-014-0192-y}
}

@article{casting-euler-les,
author = {Liu, Zhongqiu and Li, Baokuan},
year = {2017},
month = {03},
pages = {},
title = {Large-Eddy Simulation of Transient Horizontal Gas–Liquid Flow in Continuous Casting Using Dynamic Subgrid-Scale Model},
journal = {Metallurgical and Materials Transactions B},
doi = {10.1007/s11663-017-0947-3}
}

@article{casting-lagrange-bubbles,
author = {Yang, Weidong and Luo, Zhiguo and Zhao, Nannan and Zou, Zongshu},
year = {2020},
month = {08},
pages = {1160},
title = {Numerical Analysis of Effect of Initial Bubble Size on Captured Bubble Distribution in Steel Continuous Casting Using Euler-Lagrange Approach Considering Bubble Coalescence and Breakup},
volume = {10},
journal = {Metals},
doi = {10.3390/met10091160}
}

@article{casting-new-collective-dynamics-models,
  title={Numerical Analysis of Effect of Operation Conditions on Bubble Distribution in Steel Continuous Casting Mold with Advanced Bubble Break-up and Coalescence Models},
  author={Weidong Yang and Zhiguo Luo and Yingjie Gu and Zhiyuan Liu and Zongshu Zou},
  journal={ISIJ International},
  volume={Adv. Pub.},
  number={ },
  pages={},
  year={2020},
  doi={10.2355/isijinternational.ISIJINT-2020-106}
}

@article{embr-experiment,
author = {Schurmann, Dennis and Glavinic, Ivan and Willers, Bernd and Timmel, Klaus},
year = {2019},
month = {11},
pages = {},
title = {Impact of the Electromagnetic Brake Position on the Flow Structure in a Slab Continuous Casting Mold: An Experimental Parameter Study},
volume = {51},
journal = {Metallurgical and Materials Transactions B},
doi = {10.1007/s11663-019-01721-x}
}

@article{prl-path-instability,
author = {Mougin, Guillaume and Magnaudet, Jacques},
year = {2002},
month = {02},
pages = {014502},
title = {Path Instability of a Rising Bubble},
volume = {88},
journal = {Physical Review Letters},
doi = {10.1103/PhysRevLett.88.014502}
}

@article{natcomms-shape-dynamics,
author = {Tripathi, Manoj and Sahu, Kirti and Govindarajan, Rama},
year = {2015},
month = {02},
pages = {6268},
title = {Dynamics of an initially spherical bubble rising in quiescent liquid},
volume = {6},
journal = {Nature Communications},
doi = {10.1038/ncomms7268}
}

@article{spiral-to-zigzag-explained,
author = {Zhang, Jie and Ni, Ming-Jiu},
year = {2017},
month = {10},
pages = {353-373},
title = {What happens to the vortex structures when the rising bubble transits from zigzag to spiral?},
volume = {828},
journal = {Journal of Fluid Mechanics},
doi = {10.1017/jfm.2017.514}
}

@article{spiral-to-zigzag-explained-2,
author = {Zhang, Jie and Sahu, Kirti and Ni, Ming-Jiu},
year = {2020},
month = {12},
pages = {103551},
title = {Transition of bubble motion from spiralling to zigzagging: A wake-controlled mechanism with a transverse magnetic field},
volume = {136},
journal = {International Journal of Multiphase Flow},
doi = {10.1016/j.ijmultiphaseflow.2020.103551}
}

@article{shape-and-wake-simulations,
author = {Gaudlitz, Daniel and Adams, Nikolaus},
year = {2009},
month = {12},
pages = {},
title = {Numerical investigation of rising bubble wake and shape variations},
volume = {21},
journal = {Physics of Fluids},
doi = {10.1063/1.3271146}
}

@article{dns-longitudinal-field,
author = {Schwarz, S. and Fröhlich, Jochen},
year = {2014},
month = {06},
pages = {134-151},
title = {Numerical study of single bubble motion in liquid metal exposed to a longitudinal magnetic field},
volume = {62},
journal = {International Journal of Multiphase Flow},
doi = {10.1016/j.ijmultiphaseflow.2014.02.012}
}

@article{imb-transverse-field,
author = {Jin, K. and Kumar, Purushotam and Vanka, S. and Thomas, Brian},
year = {2016},
month = {09},
pages = {093301},
title = {Rise of an argon bubble in liquid steel in the presence of a transverse magnetic field},
volume = {28},
journal = {Physics of Fluids},
doi = {10.1063/1.4961561}
}

@article{zhang-mf-vertical,
author = {Zhang,Jie  and Ni,Ming-Jiu },
title = {Direct simulation of single bubble motion under vertical magnetic field: Paths and wakes},
journal = {Physics of Fluids},
volume = {26},
number = {10},
pages = {102102},
year = {2014},
doi = {10.1063/1.4896775},
URL = {https://doi.org/10.1063/1.4896775}
}

@thesis{zhang-thesis,
author = {Zhang, Chaojie},
year = {2009},
month = {04},
pages = {},
title = {Liquid metal flows driven by gas bubbles in a static magnetic field},
school = {Technischen Universität Dresden}
}

@article{zhang-mf-simulations,
author = {Zhang, Jie and Ni, Ming-Jiu and Moreau, René},
year = {2016},
month = {03},
pages = {032101},
title = {Rising motion of a single bubble through a liquid metal in the presence of a horizontal magnetic field},
volume = {28},
journal = {Physics of Fluids},
doi = {10.1063/1.4942014}
}

@article{uttt-path-instability, 
title={Force measurements on rising bubbles}, 
volume={569}, 
DOI={10.1017/S0022112006002928}, 
journal={Journal of Fluid Mechanics}, 
publisher={Cambridge University Press}, 
author={Shew, Woodrow l. and Poncet, Sebastien and Pinton, Jean-François}, 
year={2006}, 
pages={51–60}
}

@article{udv-review-article,
author = {Strumpf, Erik},
year = {2017},
month = {08},
pages = {168-185},
title = {Experimental study on rise velocities of single bubbles in liquid metal under the influence of strong horizontal magnetic fields in a flat vessel},
volume = {97},
journal = {International Journal of Multiphase Flow},
doi = {10.1016/j.ijmultiphaseflow.2017.08.001}
}

@article{uttt-x-ray-single-bubble,
author = {Richter, Thomas and Keplinger, Olga and Shevchenko, Natalia and Wondrak, T. and Eckert, Kerstin and Eckert, S. and Odenbach, S.},
year = {2018},
month = {03},
pages = {32-41},
title = {Single bubble rise in GaInSn in a horizontal magnetic field},
journal = {International Journal of Multiphase Flow},
volume = {104},
doi = {10.1016/j.ijmultiphaseflow.2018.03.012}
}

@article{udv-longitudinal-field,
author = {Zhang, Chaojie and Eckert, S. and Gerbeth, Gunter},
year = {2005},
month = {07},
pages = {824-842},
title = {Experimental study of single bubble motion in a liquid metal column exposed to a {DC} magnetic field},
volume = {31},
journal = {International Journal of Multiphase Flow},
doi = {10.1016/j.ijmultiphaseflow.2005.05.001}
}

@article{udv-transverse-field,
author = {Wang, Zenghui and Wang, S.D. and Meng, X. and Ni, M.J.},
year = {2017},
month = {05},
pages = {201-208},
title = {UDV measurements of single bubble rising in a liquid metal Galinstan with a transverse magnetic field},
volume = {94},
journal = {International Journal of Multiphase Flow},
doi = {10.1016/j.ijmultiphaseflow.2017.05.001}
}

@article{x-ray-bubble-chain-simulate,
author = {Liu, Liu and Keplinger, Olga and Ziegenhein, Thomas and Shevchenko, Natalia and Eckert, Sven and Yan, Hongjie and Lucas, Dirk},
year = {2018},
month = {09},
pages = {218-237},
title = {Euler-Euler modeling and {X-ray} measurement of oscillating bubble chain in liquid metals},
journal = {International Journal of Multiphase Flow},
volume = {110},
doi = {10.1016/j.ijmultiphaseflow.2018.09.011}
}

@article{x-ray-prime-code,
author = {Krull, Benjamin and Strumpf, E and Keplinger, Olga and Shevchenko, Natalia and Fröhlich, Jochen and Eckert, S and Gerbeth, Gunter},
year = {2017},
month = {07},
pages = {012006},
title = {Combined experimental and numerical analysis of a bubbly liquid metal flow},
volume = {228},
journal = {IOP Conference Series: Materials Science and Engineering},
doi = {10.1088/1757-899X/228/1/012006}
}

@article{x-ray-bubble-breakup,
author = {Keplinger, Olga and Shevchenko, Natalia and Eckert, S.},
year = {2019},
month = {04},
pages = {39-50},
title = {Experimental investigation of bubble breakup in bubble chains rising in a liquid metal},
volume = {116},
journal = {International Journal of Multiphase Flow},
doi = {10.1016/j.ijmultiphaseflow.2019.03.027}
}

@article{x-ray-bubble-coalescence,
author = {Keplinger, Olga and Shevchenko, Natalia and Eckert, S.},
year = {2018},
month = {04},
pages = {159-169},
title = {Visualization of bubble coalescence in bubble chains rising in a liquid metal},
journal = {International Journal of Multiphase Flow},
volume = {105},
doi = {10.1016/j.ijmultiphaseflow.2018.04.001}
}

@article{x-ray-validation,
author = {Keplinger, Olga and Shevchenko, Natalia and Eckert, S},
year = {2017},
month = {07},
pages = {012009},
title = {Validation of {X-ray} radiography for characterization of gas bubbles in liquid metals},
volume = {228},
journal = {IOP Conference Series: Materials Science and Engineering},
doi = {10.1088/1757-899X/228/1/012009}
}

@thesis{hzdr-ibm-bubbles-thesis,
	title = {An immersed boundary method for particles and bubbles in magnetohydrodynamic flows},
	type = {phdthesis},
	year = {2014},
	author = {Schwarz, Stephan},
	url = {https://nbn-resolving.org/urn:nbn:de:bsz:14-qucosa-142500}
}

@article{limmcast,
author = {Timmel, Klaus and Eckert, Sven and Gerbeth, Gunter and Stefani, Frank and Wondrak, Thomas},
year = {2010},
month = {01},
pages = {1134-1141},
title = {Experimental Modeling of the Continuous Casting Process of Steel Using Low Melting Point Metal Alloys — the LIMMCAST Program},
volume = {50},
journal = {Isij International - ISIJ INT},
doi = {10.2355/isijinternational.50.1134}
}

@article{embr-experiment-2,
author = {Thomas, Brian and Singh, Ramnik and Vanka, S. and Timmel, Klaus and Eckert, S. and Gerbeth, Gunter},
year = {2015},
month = {01},
pages = {93-104},
title = {Effect of Single-Ruler Electromagnetic Braking (EMBr) Location on Transient Flow in Continuous Casting},
volume = {15},
journal = {Journal for Manufacturing Science and Production},
doi = {10.1515/jmsp-2014-0047}
}

@article{embr-visualized,
author = {Timmel, Klaus and Shevchenko, Natalia and Röder, Michael and Anderhuber, Marc and Gardin, Pascal and Eckert, Sven and Gerbeth, Gunter},
year = {2015},
month = {04},
pages = {700-710},
title = {Visualization of Liquid Metal Two-phase Flows in a Physical Model of the Continuous Casting Process of Steel},
volume = {46},
journal = {Metallurgical and Materials Transactions B},
doi = {10.1007/s11663-014-0231-8}
}

@article{embr-cift,
author = {Wondrak, Thomas and Eckert, Sven and Gerbeth, Gunter and Klotsche, Konrad and Stefani, Frank and Timmel, Klaus and Peyton, A.J. and Terzija, Nataša and Yin, Wuliang},
year = {2011},
month = {12},
pages = {1201-1210},
title = {Combined Electromagnetic Tomography for Determining Two-phase Flow Characteristics in the Submerged Entry Nozzle and in the Mold of a Continuous Casting Model},
volume = {42},
journal = {Metallurgical and Materials Transactions B},
doi = {10.1007/s11663-011-9553-y}
}

@article{optics-collective-dynamics,
author = {Ziegenhein, Thomas and Lucas, D.},
year = {2017},
month = {03},
pages = {},
title = {Observations on bubble shapes in bubble columns under different flow conditions},
volume = {85},
journal = {Experimental Thermal and Fluid Science},
doi = {10.1016/j.expthermflusci.2017.03.009}
}

@article{saito-neutrons-1,
title = "Measurements of liquid–metal two-phase flow by using neutron radiography and electrical conductivity probe",
journal = "Experimental Thermal and Fluid Science",
volume = "29",
number = "3",
pages = "323-330",
year = "2005",
note = "Third European-Japanese Two-Phase Flow Group Meeting",
issn = "0894-1777",
doi = "https://doi.org/10.1016/j.expthermflusci.2004.05.009",
url = "http://www.sciencedirect.com/science/article/pii/S0894177704000676"
}

@article{saito-neutrons-2,
title = "Application of high frame-rate neutron radiography to liquid-metal two-phase flow research",
journal = "Nuclear Instruments and Methods in Physics Research Section A: Accelerators, Spectrometers, Detectors and Associated Equipment",
volume = "542",
number = "1",
pages = "168-174",
year = "2005",
note = "Proceedings of the Fifth International Topical Meeting on Neutron Radiography",
issn = "0168-9002",
doi = "https://doi.org/10.1016/j.nima.2005.01.095",
url = "http://www.sciencedirect.com/science/article/pii/S016890020500152X"
}

@article{megumi-x-rays,
author = {Akashi, Megumi and Keplinger, Olga and Shevchenko, Natalia and Anders, Sten and Reuter, Markus},
year = {2019},
month = {10},
pages = {},
title = {X-ray Radioscopic Visualization of Bubbly Flows Injected Through a Top Submerged Lance into a Liquid Metal},
volume = {51},
journal = {Metallurgical and Materials Transactions B},
doi = {10.1007/s11663-019-01720-y}
}

@article{neutrons-particles-stirrer-scepanskis,
author = {Sarma, Martins and Ščepanskis, Mihails and Jakovics, Andris and Thomsen, Knud and Nikoluškins, Raimonds and Vontobel, Peter and Beinerts, Toms and Bojarevics, Andris and Platacis, Erik},
year = {2015},
month = {09},
pages = {457-463},
title = {Neutron Radiography Visualization of Solid Particles in Stirring Liquid Metal},
volume = {69},
journal = {Physics Procedia},
doi = {10.1016/j.phpro.2015.07.064}
}

@article{neutrons-particles-stirrer-scepanskis-2,
author = {Ščepanskis, Mihails and Sarma, Martins and Vontobel, Peter and Trtik, Pavel and Thomsen, Knud and Jakovics, Andris and Beinerts, Toms},
year = {2017},
month = {04},
pages = {1045-1054},
title = {Assessment of Electromagnetic Stirrer Agitated Liquid Metal Flows by Dynamic Neutron Radiography},
volume = {48},
journal = {Metallurgical and Materials Transactions B},
doi = {10.1007/s11663-016-0902-8}
}

@article{neutrons-simulations-stirrer-valters,
author = {Dzelme, Valters and Jakovics, Andris and Vencels, Juris and Köppen, D. and Baake, E.},
year = {2018},
month = {10},
pages = {012047},
title = {Numerical and experimental study of liquid metal stirring by rotating permanent magnets},
volume = {424},
journal = {IOP Conference Series: Materials Science and Engineering},
doi = {10.1088/1757-899X/424/1/012047}
}

@article{gaudlitz-shape-wake-variations-1-bubble,
author = {Gaudlitz, Daniel and Adams, Nikolaus},
year = {2009},
month = {12},
pages = {},
title = {Numerical investigation of rising bubble wake and shape variations},
volume = {21},
journal = {Physics of Fluids},
doi = {10.1063/1.3271146}
}

@article{taborda-les-euler-lagrange,
author = {Taborda, Manuel and Sommerfeld, Martin and Muniz, Marcelo},
year = {2021},
month = {01},
pages = {116121},
title = {LES-Euler/Lagrange modelling of bubble columns considering mass transfer, chemical reactions and effects of bubble dynamics},
volume = {229},
journal = {Chemical Engineering Science},
doi = {10.1016/j.ces.2020.116121}
}

@article{in-depth-study-of-ellipsoid-kinematics,
author = {Will, Jelle and Mathai, Varghese and Huisman, Sander and Lohse, D. and Sun, Chao and Krug, Dominik},
year = {2021},
month = {04},
pages = {A16},
title = {Kinematics and dynamics of freely rising spheroids at high Reynolds numbers},
volume = {912},
journal = {Journal of Fluid Mechanics},
doi = {10.1017/jfm.2020.1104}
}

@article{hele-shaw-bubbles-vof,
author = {Wang, Xue and Klaasen, Bart and Degrève, Jan and Mahulkar, Amit and Heynderickx, Geraldine and Reyniers, Marie-Françoise and Blanpain, Bart and Verhaeghe, Frederik},
year = {2016},
month = {05},
pages = {053304},
title = {Volume-of-fluid simulations of bubble dynamics in a vertical Hele-Shaw cell},
volume = {28},
journal = {Physics of Fluids},
doi = {10.1063/1.4948931}
}

@article{hele-shaw-bubbles-experiment,
author = {Roig, Veronique and Roudet, Matthieu and Risso, Frédéric and Billet, Anne-Marie},
year = {2012},
month = {09},
pages = {444-466},
title = {Dynamics of a high-Reynolds-number bubble rising within a thin gap},
volume = {707},
journal = {Journal of Fluid Mechanics},
doi = {10.1017/jfm.2012.289}
}

@article{sommer-4d-ptv,
author = {Sommer, A.-E. and Nikpay, Mitra and Heitkam, Sascha and Rudolph, Martin and Eckert, Sven},
year = {2018},
month = {08},
pages = {116–122},
title = {A novel method for measuring flotation recovery by means of 4D particle tracking velocimetry},
volume = {124},
journal = {Minerals Engineering},
doi = {10.1016/j.mineng.2018.05.006}
}

@article{sommer-pept,
author = {Sommer, A.-E and Ortmann, K. and van Heerden, Michael and Richter, Thomas and Leadbeater, Thomas and Cole, Katie and Heitkam, S. and Brito-Parada, P.R. and Eckert, Kerstin},
year = {2020},
month = {09},
pages = {106410},
title = {Application of Positron Emission Particle Tracking (PEPT) to measure the bubble-particle interaction in a turbulent and dense flow},
volume = {156},
journal = {Minerals Engineering},
doi = {10.1016/j.mineng.2020.106410}
}

@article{neutrons-fluid-flow-visuals,
author = {Cimbala, John and Hughes, Daniel and Levine, Samuel and Sathianathan, Dhushy},
year = {1988},
month = {06},
pages = {},
title = {Application of Neutron Radiography for Fluid Flow Visualization},
volume = {81:3},
journal = {Nucl. Technol.; (United States)},
doi = {10.13182/NT88-A16065}
}

@article{particle-boundary-conditions,
author = {May, Ronja and Gruy, Frédéric and Fröhlich, Jochen},
year = {2018},
month = {12},
pages = {1-2},
title = {Impact of particle boundary conditions on the collision rates of inclusions around a single bubble rising in liquid metal},
volume = {18},
journal = {PAMM},
doi = {10.1002/pamm.201800029}
}

@article{udv-wake-flow-structure,
author = {Zhang, Chaojie and Eckert, Sven and Gerbeth, Gunter},
year = {2007},
month = {03},
pages = {57 - 82},
title = {The flow structure of a bubble-driven liquid-metal jet in a horizontal magnetic field},
volume = {575},
journal = {Journal of Fluid Mechanics},
doi = {10.1017/S0022112006004423}
}

@article{metal-strring-1,
author = {Lou, Wentao and Zhu, Miaoyong},
year = {2014},
month = {10},
pages = {1706-1722},
title = {Numerical Simulation of Desulfurization Behavior in Gas-Stirred Systems Based on Computation Fluid Dynamics–Simultaneous Reaction Model (CFD–SRM) Coupled Model},
volume = {45},
journal = {Metallurgical and Materials Transactions B},
doi = {10.1007/s11663-014-0105-0}
}

@article{metal-strring-2,
author = {Liu, Yu and Ersson, Mikael and Liu, Heping and Jönsson, Pär and Gan, Yong},
year = {2018},
month = {11},
pages = {},
title = {A Review of Physical and Numerical Approaches for the Study of Gas Stirring in Ladle Metallurgy},
volume = {50},
journal = {Metallurgical and Materials Transactions B},
doi = {10.1007/s11663-018-1446-x}
}

@article{metal-strring-3,
author = {Cao, Qing and Nastac, Laurentiu},
year = {2018},
month = {01},
pages = {1-8},
title = {Numerical modelling of the transport and removal of inclusions in an industrial gas-stirred ladle},
volume = {45},
journal = {Ironmaking \& Steelmaking},
doi = {10.1080/03019233.2018.1426697}
}

@article{metal-strring-4,
author = {Morales, Rodolfo and Calderon Hurtado, Fabian Andres and Chattopadhyay, Kinnor and Guarneros, Sergio},
year = {2020},
month = {01},
pages = {},
title = {Physical and Mathematical Modeling of Flow Structures of Liquid Steel in Ladle Stirring Operations},
volume = {51},
journal = {Metallurgical and Materials Transactions B},
doi = {10.1007/s11663-019-01759-x}
}

@article{metal-strring-5,
author = {Ramasetti, Eshwar and Visuri, Ville-Valtteri and Sulasalmi, Petri and Palovaara, Tuomas and Gupta, Avishek and Fabritius, Timo},
year = {2019},
month = {06},
pages = {},
title = {Physical and CFD Modeling of the Effect of Top Layer Properties on the Formation of Open‐Eye in Gas‐Stirred Ladles With Single and Dual‐Plugs},
volume = {90},
journal = {steel research international},
doi = {10.1002/srin.201900088}
}

@inbook{flotation-book-1,
author = {Nguyen, Anh and Schulze, and},
year = {2004},
month = {01},
pages = {},
title = {Colloidal Science of Flotation},
isbn = {0-8247-4782-8},
doi = {10.1201/9781482276411}
}

@inbook{flotation-book-2,
author = {Fuerstenau, M.C. and Yoon, R.-H and Jameson, G.J.},
year = {2007},
month = {01},
pages = {},
title = {Froth flotation: A century of innovation},
volume = {1},
publisher = {Society for Mining, Metallurgy, and Exploration, Littleton, Colo., Ix},
isbn = {978-0-87335-280-2}
}

@article{pept-1,
author = {Burnard, David and Caden, A.J. and Gargiuli, Joseph and Leadbeater, Thomas and Parker, D. and Griffiths, William},
year = {2014},
month = {05},
pages = {43-48},
title = {A Positron Emission Particle Tracking (PEPT) Study of Inclusions in Liquid Aluminium Alloy},
volume = {922},
journal = {Advanced Materials Research},
doi = {10.4028/www.scientific.net/AMR.922.43}
}

@inbook{pept-2,
author = {Burnard, David and Gargiuli, Joseph and Leadbeater, Thomas and Parker, D. and Griffiths, William},
year = {2011},
month = {06},
pages = {25-28},
title = {The Application of Positron Emission Particle Tracking (PEPT) to Study Inclusions in the Casting Process},
volume = {690},
journal = {Materials Science Forum},
doi = {10.4028/www.scientific.net/MSF.690.25}
}

@article{pept-3,
author = {Griffiths, William and Gaber Beshay, Youssef and Caden, A. and Fan, X. and Gargiuli, Joseph and Leadbeater, Thomas and Parker, D.},
year = {2011},
month = {04},
pages = {},
title = {The Use of Positron Emission Particle Tracking (PEPT) to Study the Movement of Inclusions in Low-Melting-Point Alloy Castings},
volume = {43},
journal = {Metallurgical and Materials Transactions B},
doi = {10.1007/s11663-011-9596-0}
}

@article{pept-4,
author = {Dybalska, Agnieszka and Caden, Adrian and Parker, D. and Wedderburn, John and Griffiths, William},
year = {2020},
month = {07},
pages = {},
title = {Liquid Metal Flow Studied by Positron Emission Tracking},
volume = {51},
journal = {Metallurgical and Materials Transactions B},
doi = {10.1007/s11663-020-01897-7}
}

@article{klevs2021dynamic,
author = {Klevs,M.  and Birjukovs,M.  and Zvejnieks,P.  and Jakovics,A. },
title = {Dynamic mode decomposition of magnetohydrodynamic bubble chain flow in a rectangular vessel},
journal = {Physics of Fluids},
volume = {33},
number = {8},
pages = {083316},
year = {2021},
doi = {10.1063/5.0054831}
}

@book{clift-bubbles,
author = {Clift, Roland and Grace, J and Weber, M},
year = {1978},
month = {01},
pages = {},
title = {Bubbles, Drops, and Particles},
isbn = {9780121769505}
}

@inproceedings{non-local-means-filter,
author = {Coll, Bartomeu and Morel, Jean-Michel},
year = {2005},
month = {07},
pages = {60- 65 vol. 2},
title = {A non-local algorithm for image denoising},
volume = {2},
isbn = {0-7695-2372-2},
journal = {Proceedings of the IEEE Computer Society Conference on Computer Vision and Pattern Recognition},
doi = {10.1109/CVPR.2005.38}
}

@article{non-local-means-filter-new-weight-function,
author = {Coll, Bartomeu and Morel, Jean-Michel},
year = {2011},
month = {09},
pages = {},
title = {Non-Local Means Denoising},
volume = {1},
journal = {Image Processing On Line},
doi = {10.5201/ipol.2011.bcm_nlm}
}

@article{local-adaptive-thresholding,
author = {Sezgin, M. and Sankur, Bulent},
year = {2004},
month = {01},
pages = {},
title = {Comparison of thresholding methods for non-destructive testing applications}
}

@article{images-mathematical-morphology,
author = {Haralick, Robert and Sternberg, Stanley and Zhuang, Xinhua},
year = {1987},
month = {08},
pages = {532 - 550},
title = {Image Analysis Using Mathematical Morphology},
volume = {PAMI-9},
journal = {Pattern Analysis and Machine Intelligence, IEEE Transactions on},
doi = {10.1109/TPAMI.1987.4767941}
}

@article{otsu-thresholding,
author = {Otsu, Nobuyuki},
year = {1979},
month = {01},
pages = {62-66},
title = {A Threshold Selection Method from Gray-Level Histograms},
volume = {9},
journal = {Systems, Man and Cybernetics, IEEE Transactions on}
}

@online{wolfram-mathematica-inpaint, 
organization={Wolfram Research}, 
title={Inpaint}, 
year={2015}, 
url={https://reference.wolfram.com/language/ref/Inpaint.html}
}

@inbook{x-ray-neutron-experiments-book,
author = {Lappan, Tobias and Sarma, Martins and Heitkam, Sascha and Mannes, David and Trtik, Pavel and Shevchenko, Natalia and Eckert, Kerstin and Eckert, Sven},
year = {2021},
month = {02},
pages = {13-29},
title = {X-Ray and Neutron Radiographic Experiments on Particle-Laden Molten Metal Flows},
isbn = {978-3-030-65252-4},
doi = {10.1007/978-3-030-65253-1_2}
}

@article{heitkam-particles-froth-2019,
author = {Heitkam, Sascha and Lappan, Tobias and Eckert, Sven and Trtik, Pavel and Eckert, Kerstin},
year = {2019},
month = {04},
pages = {},
title = {Tracking of Particles in Froth Using Neutron Imaging},
volume = {91},
journal = {Chemie Ingenieur Technik},
doi = {10.1002/cite.201800127}
}

@article{2021-review-article-bubbles-in-liquid-metal,
author = {Haas, Tim and Schubert, Christian and Eickhoff, Moritz and Pfeifer, Herbert},
year = {2021},
month = {04},
pages = {664},
title = {A Review of Bubble Dynamics in Liquid Metals},
volume = {11},
journal = {Metals},
doi = {10.3390/met11040664}
}

@article{sten-spectral-random-masking,
author = {Anders, Sten and Noto, Daisuke and Seilmayer, Martin and Eckert, Sven},
year = {2019},
month = {03},
pages = {68},
title = {Spectral random masking: a novel dynamic masking technique for PIV in multiphase flows},
volume = {60},
journal = {Experiments in Fluids},
doi = {10.1007/s00348-019-2703-8}
}

@article{sten-srm-velocity-temp-measurements,
author = {Anders, Sten and Noto, Daisuke and Tasaka, Yuji and Eckert, Sven},
year = {2020},
month = {04},
pages = {},
title = {Simultaneous optical measurement of temperature and velocity fields in solidifying liquids},
volume = {61},
journal = {Experiments in Fluids},
doi = {10.1007/s00348-020-2939-3}
}

@misc{birjukovs2021resolving,
      title={Resolving gas bubbles ascending in liquid metal from low-SNR neutron radiography images}, 
      author={Mihails Birjukovs and Pavel Trtik and Anders Kaestner and Jan Hovind and Martins Klevs and Knud Thomsen and Andris Jakovics},
      year={2021},
      eprint={2109.04883},
      archivePrefix={arXiv},
      primaryClass={physics.flu-dyn}
}

\end{document}